%% file: root.tex
\documentclass[twocolumn]{./docstyle/autart}
\input{./docstyle/defs.tex}

\begin{document}
\begin{frontmatter}
    \title{Stability of Convex Time-Varying\\Receding-Horizon Potential Games}
    \thanks[footnoteinfo]{Corresponding author.}
    
    \author[TUHH]{Alex Erdin\thanksref{footnoteinfo}}\ead{alex.erdin@tuhh.de},
    \author[ETH]{Sophie Hall}\ead{shall@ethz.ch},
    \author[TUHH]{Timm Faulwasser}\ead{timm.faulwasser@ieee.org}

    \address[ETH]{Automatic Control Laboratory, ETH Zurich, Switzerland}
    \address[TUHH]{Institute of Control Systems, Hamburg University of Technology, Germany}
    
    
    \begin{abstract} 
        \Acp{rhg} are an emerging control methodology applying open-loop equilibrium solutions of competitive games in repeating fashion to tractably solve dynamic games over long horizons.
        In this paper, we extend recent dissipativity-based stability results for \acp{rhg} to a time-varying setting, where the optimal operating behaviour is characterised by a time-varying \ac{gne} rather than a steady-state one. To this end, we build upon a connection between \acp{rhg} and \ac{empc}, which allows us to derive $\mathbb{P}$-practical asymptotic stability with respect to the optimal (\ac{gne}) trajectory. Our results hold for time-varying, quadratic objectives, \ac{ltv} dynamics and constant constraints in the discrete-time setting and we draw upon a numerical example to illustrate our findings.
    \end{abstract}
\end{frontmatter}
\date{24 August 2026}

\input{sections/01_introduction}
\input{sections/02_rhgs}
\input{sections/03_examples}
\input{sections/98_conclusion}
\input{sections/99_appendix}

\renewcommand*{\bibfont}{\scriptsize}
\printbibliography[heading=bibintoc]
\end{document}

%% file: docstyle/defs.tex
\usepackage[utf8]{inputenc}
\usepackage{amsmath, amsfonts, amssymb, bbm, mathrsfs, accents}
\usepackage[acronym,shortcuts]{glossaries}
\usepackage{enumitem}
\usepackage{mdframed}
\usepackage{booktabs}
\usepackage{multirow}
\usepackage{array}
\usepackage{url}
\usepackage{graphicx}
\usepackage[dvipsnames]{xcolor}
\usepackage{placeins}
\usepackage{tikz}
\usetikzlibrary{arrows.meta, positioning, fit, calc}

\usepackage[hidelinks]{hyperref}


\usepackage[%
    style=authoryear,
    maxcitenames=3,
    uniquename=false,
    dashed=false,
    doi=false,
    isbn=false,
    url=false,
    eprint=false,
    backend=bibtex,
    ]{biblatex}
\DeclareDelimFormat{finalnamedelim}{\addnbspace\&\space}

\AtEveryBibitem{%
    \clearfield{note}%
    \clearfield{day}%
    \clearfield{month}%
}

\definecolor{autblue}{RGB}{0, 128, 172}
\AtEveryCite{\color{autblue}}

\definecolor{ibmblue}{HTML}{648FFF}
\definecolor{ibmpurple}{HTML}{785EF0}
\definecolor{ibmpink}{HTML}{DC267F}
\definecolor{ibmgreen}{HTML}{009E73}
\definecolor{ibmorange}{HTML}{FE6100}
\definecolor{ibmyellow}{HTML}{FFB000}

\newdimen\mgn\mgn=5pt
\tikzstyle{box} = [draw,
                   inner sep=0pt,
                   rounded corners=\mgn,
                   align=center,
                   font=\small]
\tikzstyle{labelStyle} = [anchor=south west, 
                          outer sep=0pt,
                          inner sep=0pt,
                          yshift=\mgn,
                          font=\normalfont]
\tikzstyle{cornerBadge} = [anchor=east, 
                          outer sep=0pt,
                          inner sep=1pt,
                          xshift=-\mgn,
                          rounded corners=2pt,
                          font=\color{white}\bf\sffamily\tiny]                 
\tikzstyle{arrow} = [-latex, shorten <=-\mgn, shorten >=-\mgn]

\tikzset{
    hyperlink node/.style n args={3}{
        alias=sourcenode,
        /utils/exec={%
            \xdef\HLxsep{\pgfkeysvalueof{/pgf/outer xsep}}%
            \xdef\HLysep{\pgfkeysvalueof{/pgf/outer ysep}}%
        },
        append after command={
            let \p1=([shift={(\HLxsep, -\HLysep)}] sourcenode.north west),
                \p2=([shift={(-\HLxsep, \HLysep)}] sourcenode.south east),
                \n1={\x2-\x1},
                \n2={\y1-\y2} in
            node [anchor=north west, at=(\p1), inner sep=0pt, outer sep=0pt] (last) {\expandafter\hyperref\expandafter[#1]{\XeTeXLinkBox{\phantom{\rule{\n1}{\n2}}}}}
            node[anchor=east, outer sep=0pt, inner sep=1pt, xshift=-5pt, rounded corners=2pt, font=\color{white}\bf\sffamily\tiny, fill=#3] at (last.north east) {#2 \ref{#1}}
        }
    },
}

\newacronym{ltv}{LTV}{linear time-varying}
\newacronym{ocp}{OCP}{optimal control problem}
\newacronym{pg}{PG}{potential game}
\newacronym{gpg}{GPG}{generalized potential game}
\newacronym[longplural={generalized Nash equilibria}]{gne}{GNE}{generalized Nash equilibrium}
\newacronym{vgne}{v-GNE}{variational GNE}
\newacronym{nep}{NEP}{Nash equilibrium problem}
\newacronym{gnep}{GNEP}{generalized Nash equilibrium problem}
\newacronym{empc}{EMPC}{economic model predictive control}
\newacronym{mpc}{MPC}{model predictive control}
\newacronym{kkt}{KKT}{Karush-Kuhn-Tucker}
\newacronym{rhg}{RHG}{receding-horizon game}

\newcommand{\N}{\ensuremath{\mathbb{N}}} 
\newcommand{\bx}{\ensuremath{\boldsymbol{x}}} 
\newcommand{\bu}{\ensuremath{\boldsymbol{u}}} 
\newcommand{\bz}{\ensuremath{\boldsymbol{z}}} 
\newcommand{\X}{\ensuremath{\mathbbm{X}}} 
\newcommand{\bX}{\ensuremath{\pmb{\mathbbm{X}}}} 
\newcommand{\U}{\ensuremath{\mathbbm{U}}} 
\newcommand{\bU}{\ensuremath{\pmb{\mathbbm{U}}}} 
\newcommand{\V}{\ensuremath{\mathcal{V}}} 
\newcommand{\domain}[1]{\ensuremath{\mathbb{R}^{#1}}} 
\newcommand{\partderiv}[2]{\frac{\partial\, #1}{\partial\, #2}} 
\newcommand{\partderivv}[2]{\frac{\partial^2\, #1}{\partial\, #2^2}} 
\DeclareMathOperator{\col}{col} 
\DeclareMathOperator{\blkdiag}{blkdiag} 
\DeclareMathSymbol{\shortminus}{\mathbin}{AMSa}{"39} 
\newcommand{\lowtilde}[1]{\accentset{\sim}{#1}} 

%% file: sections/01_introduction.tex
\section{Introduction} \label{introduction}
Game-theoretic control schemes have gained significant popularity in recent years due to their relevance in multi-agent systems. Unlike distributed optimal control, where agents jointly optimize a common objective, game-theoretic frameworks involve agents that seek to selfishly optimize their own objectives. Such game-theoretic formulations arise naturally in many robotic and sociotechnical frameworks, making a thorough theoretical underpinning essential \parencite{hall_stability_2024,bhatt_efficient_2023,zanardi_urban_2021}.

The utility of such formulations has been demonstrated across a range of applications: for instance, \cite{fisac_hierarchical_2019} and \cite{bahram_game-theoretic_2016} show that explicitly modelling competitive interactions can resolve the frozen robot problem \parencite{trautman_unfreezing_2010}, while \cite{cenedese_highway_2021} demonstrate that game-theoretic incentive design enables indirect control of systems without resorting to costly hard constraints.

Underlying many of these approaches is a deeper structural connection: \cite{marden_connections_2007} show that many distributed control approaches including formation control, consensus problems and cooperative control \parencite{jadbabaie_coordination_2003,olfati-saber_consensus_2004,olfati-saber_consensus_2007} can be reformulated in game-theoretic terms, connecting them to \acp{pg} as formalized by \cite{monderer_potential_1996}. This connection was later consolidated by \cite{li_designing_2013} through their work on game design for distributed control which laid the groundwork for a variety of explicit game-theoretic algorithms for multi-agent control, with many adopting a potential game foundation.

\Ac{pg}-frameworks require certain problem structure, yet remain expressive enough to find application in a wide range of different systems: \cite{arslan_autonomous_2007}, propose two control schemes utilizing potential functions for the task of vehicle-target assignment. \cite{mohsenian-rad_optimal_2010,eksin_demand_2015} and \cite{zazo_dynamic_2016} formulate the problem of demand-side management as a dynamic \ac{pg}, proposing various distributed energy consumption scheduling algorithms. Relatedly, \cite{grammatico_dynamic_2017} studies dynamic aggregative games, a subclass of dynamic \acp{pg}, providing a model-free solution to battery charging. Other applications include autonomous driving \parencite{zanardi_urban_2021,evens_learning_2022}, traffic congestion games \parencite{cenedese_highway_2021} and general congestion control \parencite{govaert_relative_2019}. Spanning a wide range of problem structures and application domains, all of these results and the underlying algorithms either rely on specifically tailored convergence results or do not provide any guarantees. This, in turn, underpins the need for a general framework for stability analysis of dynamic receding horizon games.

A first step towards a general framework for stability analysis is made by \cite{cruz_moving_2001,van_den_broek_moving_2002} with the introduction of moving horizon games, drawing an analogy to \ac{mpc} \parencite{rawlings_model_2017}. The approach was later extended by \cite{hall_receding_2022}, who made the connection between \acp{rhg} and \ac{mpc} theory explicit for systems with time-invariant cost and dynamics. Many real-world (multi-agent) systems, however, are inherently time-varying: electric grids operate under dynamic pricing schemes \parencite{hall_receding_2022}, traffic networks experience time-dependent congestion patterns, and robotic systems are subject to changing environmental conditions \parencite{bhatt_efficient_2023}. Extending the stability analysis of \acp{rhg} to the time-varying setting is therefore crucial.

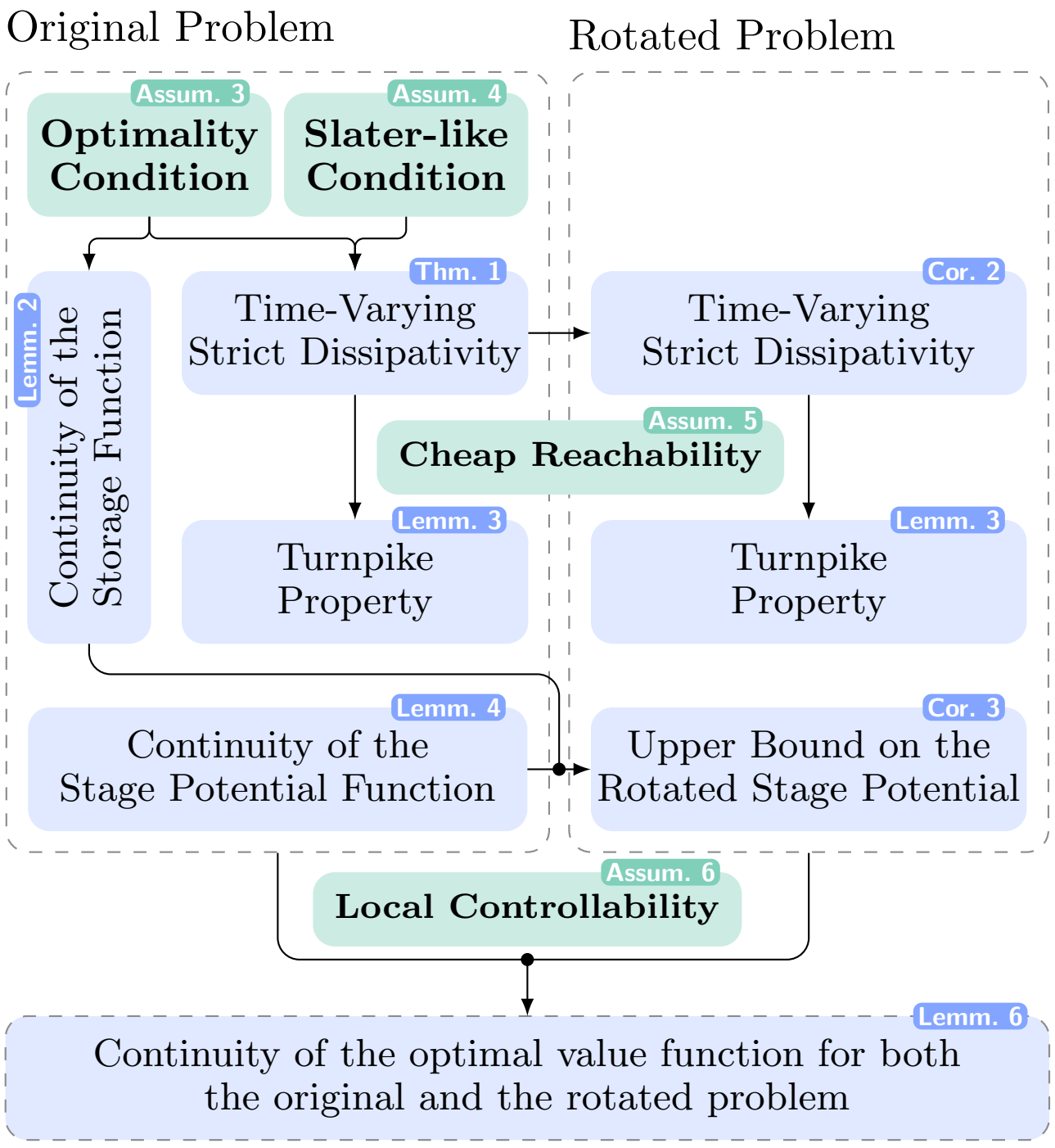
\begin{figure}[t]
    \centering
    \input{figures/assumptions}\vspace{-10pt}
    \caption{Map of our assumptions (\textcolor{ibmgreen!50}{\rule{1.5ex}{1.5ex}} green) and the ones from \cite{grune_economic_2020} (\textcolor{ibmblue!80}{\rule{1.5ex}{1.5ex}} blue).}
    \label{fig:Assumptions}
\end{figure}

This paper presents the findings of the MSc thesis \cite{erdin_stability_2025} and it addresses this gap by extending the stability results of \cite{hall_receding_2022} to the time-varying setting, where the optimal operating behaviour is characterised by a time-varying \ac{gne} rather than a steady-state one. Our analysis builds on \ac{empc} results of \cite{grune_economic_2020} that show $\mathbb{P}$-practical asymptotic stability of time-varying \ac{empc}, from which we derive a new set of sufficient conditions given our \acl{ltv} dynamics, quadratic cost and convex constraints. Most notably, we show time-varying strict dissipativity for \acp{gnep} under a \ac{pg} structure and an optimality condition. Figure \ref{fig:Assumptions} provides a graphical summary of our results and their relation to \cite{grune_economic_2020}.

The remainder of this paper is structured as follows: Section \ref{sec:Preliminaries} defines the general problem setup and introduces the multi-agent game. Section \ref{sec:PGs} presents a reformulation of this game. Section \ref{sec:Stability} analyses the stability of the game by leveraging this reformulation. Section \ref{sec:SimulationStudy} showcases the theoretical results based on an example from demand-side management. The paper ends with the conclusion in Section \ref{sec:Conclusion}.

%% file: figures/assumptions.tex
\noindent
\begin{tikzpicture}[%
    dashedBox/.style={box, minimum height=6.3cm, dashed, draw=gray!90},
    contentBox/.style={box, minimum width=2.8cm, minimum height=1cm, outer sep=\mgn, fill=ibmblue!20, dashed, draw=none}
    ]
    \newdimen\orgwidth\orgwidth=.52\linewidth
    \node[dashedBox, anchor=west, minimum width=\orgwidth, label={[labelStyle]north west:Original Problem}] (original) at (0, 0) {};
    \node[contentBox, anchor=north west, minimum width=.5\orgwidth-1.25\mgn, draw=none, fill=ibmgreen!20, hyperlink node={assum:OptimalityCondition}{Assum.}{ibmgreen!50}] (opt) at (original.north west) {\bf Optimality\\\bf Condition};
    \node[contentBox, anchor=north east, minimum width=.5\orgwidth-1.25\mgn, draw=none, fill=ibmgreen!20, hyperlink node={assum:SlaterlikeCondition}{Assum.}{ibmgreen!50}] (slater) at (original.north east) {\bf Slater-like\\\bf Condition};
    \node[contentBox, anchor=north east, inner xsep=8pt, rotate=90] (store) at ([yshift=-.5\mgn]opt.south west) {Continuity of the \\Storage Function};
    \node[contentBox, anchor=north east, hyperlink node={thm:StrictDissipativity}{Thm.}{ibmblue!80}] (diss) at ([yshift=-.5\mgn]slater.south east) {Time-Varying\\Strict Dissipativity};
    \node[contentBox, anchor=south east, hyperlink node={lem:Turnpike}{Lemm.}{ibmblue!80}] (turn) at (original.east |- store.west) {Turnpike\\Property};
    \node[contentBox, anchor=south, minimum width=\orgwidth-2\mgn, hyperlink node={lem:ContinuousStagePotential}{Lemm.}{ibmblue!80}] (stage) at (original.south) {Continuity of the\\Stage Potential Function};
    \node [anchor=north west, inner sep=0pt, outer sep=0pt] at ([xshift=\mgn,yshift=-\mgn]store.north east) {\hyperref[lem:ContinuousStorageFunction]{\XeTeXLinkBox{\phantom{\rule{30pt}{85pt}}}}};
    \node[cornerBadge, fill=ibmblue!80, rotate=90] at ([xshift=2\mgn,yshift=-2\mgn]store.north east) {Lemm. \ref{lem:ContinuousStorageFunction}};
    
    \newdimen\modwidth\modwidth=\dimexpr(\linewidth-\orgwidth-\mgn)\relax
    \node[dashedBox, anchor=east, minimum width=\modwidth, label={[labelStyle]north west:Rotated Problem}] (modified) at (\linewidth, 0) {};
    \node[contentBox, anchor=north, minimum width=\modwidth-2\mgn, hyperlink node={cor:RotatedStrictDissipativity}{Cor.}{ibmblue!80}] (diss2) at (modified.center |- diss.north) {Time-Varying\\Strict Dissipativity};
    \node[contentBox, anchor=south, minimum width=\modwidth-2\mgn, hyperlink node={lem:Turnpike}{Lemm.}{ibmblue!80}] (turn2) at (modified.center |- turn.south) {Turnpike\\Property};
    \node[contentBox, anchor=south, minimum width=\modwidth-2\mgn, hyperlink node={cor:UpperBoundedRotatedStagePotential}{Cor.}{ibmblue!80}] (above) at (modified.south) {Upper Bound on the\\Rotated Stage Potential};

    \node[contentBox, draw=gray!90, below=6.5\mgn of original.south west, anchor=north west, minimum width=\linewidth, minimum height=1cm, xshift=-\mgn, fill=ibmblue!20, hyperlink node={lem:ContinuousValueFunction}{Lemm.}{ibmblue!80}] (cont) {%
        Continuity of the optimal value function for both\\the original and the rotated problem
    };

    \draw[-latex, rounded corners=2pt] ([yshift=\mgn]opt.south) -- ++(0,-\mgn) -| ([yshift=-\mgn]store.east);
    \draw[-latex, rounded corners=2pt] ([yshift=\mgn]opt.south) -- ++(0,-\mgn) -| ([yshift=-\mgn]diss.north);
    \draw[-latex, rounded corners=2pt] ([yshift=\mgn]slater.south) -- ++(0,-\mgn) -| ([yshift=-\mgn]diss.north);
    
    \draw[arrow]
        (diss.south) edge node (cr) {} (turn.north)
        (diss2.south)edge (turn2.north)
        (diss.east)   to  (diss2.west);
    
    \draw[arrow] (stage.east) -- (above.west) coordinate[midway] (midway);
    \draw[-{Circle[length=3pt]}, shorten <=\mgn, shorten >=-1.5pt, rounded corners=\mgn] (store.west) -- ++(0,-2pt) -| (midway);

    \draw[latex-{Circle[length=3pt]},  shorten <=-\mgn, shorten >=-1.5pt]
        (cont.north) -- node (lc) {} ([yshift=8pt]cont.north) coordinate (end);
    \draw[rounded corners=\mgn] (original.south) |- (end) (modified.south) |- (end);

    \node[right, contentBox, minimum height=0.6cm, inner xsep=5pt, fill=ibmgreen!20, font=\bf\scriptsize, hyperlink node={assum:CheapReachability}{Assum.}{ibmgreen!50}] at (cr) {Cheap Reachability};
    \node[above, contentBox, minimum height=0.6cm, inner xsep=5pt, fill=ibmgreen!20, yshift=2pt, font=\bf\scriptsize, hyperlink node={assum:LocalControllability}{Assum.}{ibmgreen!50}] at (lc) {Local Controllability};
\end{tikzpicture}

%% file: sections/02_rhgs.tex
\section{Problem Statement and Preliminaries} \label{sec:Preliminaries}

Let $\V := \{1,\ldots,M\}$ be the set of all $M$ agents. For agent $v \in \V$, $x_t^v \in \domain{n^v}$ and $u_t^v \in \domain{m^v}$ denote the (closed-loop) state and input at time $t$, while $x_{k|t}^v$ and $u_{k|t}^v$ represent the open-loop state and input predictions at step $k$ relative to time $t$. We indicate stacked quantities formed by all agents using boldface letters, such as the stacked state and input vectors $\bx_t := \col(x_t^1,\ldots,x_t^M) \in \domain{n}$ and $\bu_t$, respectively. Similarly, $\bx_t^{-v}$ and $\bu_t^{-v}$ denote the stacked vectors excluding agent $v$, occasionally writing $(x_t^v,\bx_t^{-v})$ instead of $\bx_t$ to emphasize an expression's dependency on agent $v$. To denote entire trajectory pairs we write $(x^v,u^v)$ for (closed-loop) trajectories over all $t \in \N$ and $(x_{\bullet|t}^v, u_{\bullet|t}^v)$ for open-loop trajectories stacked over the entire prediction horizon $N$, respectively.

We consider a multi-agent system with $M$ self-interested agents $v$ whose \ac{ltv} dynamics are given by
\begin{equation} \label{eq:Dynamics}
    x_{t+1}^v = A_t^v x_t^v + B_t^v u_t^v,\quad \rlap{$\forall v \in \V$.}
\end{equation}
Although the agents are self-interested, their actions and states are jointly restricted through so called coupling constraints which model access to a shared resource (i.e., energy, water, etc.) or infrastructure (i.e., roads, networks, etc.).\!\footnote{Coupling constraints are sometimes also referred to as jointly convex constraints in the literature.}\! Specifically, we consider the compact, convex coupling constraints defined by
\begin{equation} \label{eq:Constraints}
    \begin{alignedat}{2}
        x_{t}^v &\in \X^v\bigl(\bx_{t}^{-v}\bigr) &&:= \bigl\{x_{t}^v \in \domain{n^v}   \mid \boldsymbol{g}_{\bx}(x_{t}^v,\bx_{t}^{-v}) \leq 0 \bigr\},\\
        u_{t}^v &\in \U^v\bigl(\bu_{t}^{-v}\bigr) &&:= \bigl\{u_{t}^v \in \domain{m^v}\! \mid \boldsymbol{g}_{\bu}(u_{t}^v,\bu_{t}^{-v\!}) \leq 0\bigr\},
    \end{alignedat}
\end{equation}
with $\boldsymbol{g}(\bx_t,\bu_t) := \col(\boldsymbol{g}_{\bx}(\bx_t),\boldsymbol{g}_{\bu}(\bu_t))$ denoting the joint inequality constraints over state and input.

Each agent $v$ aims to minimize its own time-varying quadratic stage cost
\begin{flalign} \label{eq:StageCost}
    \ell_t^v(\bx_t,\!\bu_t) \!:&\!=\! \bx_t^\top\! \boldsymbol{Q}_t^v \bx_t + \bu_t^\top\! \boldsymbol{R}_t^v \bu_t + q_t^v x_t^v + r_t^v u_t^{v\!}&&\\[3pt]
    &\!=\! \sum_{i \in \V} \!\left((x_t^v){\!}^\top\! Q_t^{v,i\!} x_t^i \!+\! (u_t^v){\!}^\top\! R_t^{v,i\!} u_t^i\right) \!\!+\! q_t^v x_t^v \!+\! r_t^v u_t^v\!,\hspace{-15pt}&&\notag
\end{flalign}
where the system matrices $(A_t^v, B_t^v)$, cost matrices $(\boldsymbol{Q}_t^v, \boldsymbol{R}_t^v)$, and cost vectors $(q_t^v, r_t^v)$ are uniformly bounded for all $t \in \N$ and $v \in \V$.

Note that this stage cost introduces a coupling among the agents in its quadratic cost terms, since they depend on the other agents decision variables $(\bx_t^{-v},\bu_t^{-v})$ as well.
Further note that without loss of generality and to avoid ambiguity, we assume $\boldsymbol{Q}_t^v$ and $\boldsymbol{R}_t^v$ are block-sparse, with only their $v$-th block row nonzero, allowing us to denote the $i$-th block matrix of agent $v$ by $Q_t^{v,i}$ and $R_t^{v,i}$, respectively.

Minimizing \eqref{eq:StageCost} over an infinite time horizon yields the following infinite-horizon \ac{gnep}
\begin{equation} \label{prob:InfGNEP} \tag{$\mathscr{P}_\infty^v$}
    \!\forall v \in \V :\left\{\rule{0pt}{44pt}\right.
    \renewcommand\arraystretch{1.4}
    \begin{array}{@{\;}r@{\quad}l@{}}\displaystyle
    \inf_{\!\!\!x^v, u^v\!\!\!} & \sum\limits_{t=0}^{\infty} \ell_t^v(\bx_t,\bu_t) - \xi_t^v \\
    \text{s.t.}   & \forall\, t \in \N \!\boldsymbol{:}\\
                  & x_{t+1}^v = A_{t}^v x_t^v + B_t^v u_t^v,\\
                  & x_t^v \in \X^v\bigl(\bx_t^{-v}\bigr),\qquad\\
                  & u_t^v \in \U^v\bigl(\bu_t^{-v}\bigr),\qquad
    \end{array}\hspace{\linewidth minus 1fil}
\end{equation}
where $\xi_t^v$ denotes some time-varying shift which will become of importance later on in guaranteeing well-posedness of Problem~\eqref{prob:InfGNEP}. Note that this shift $\xi_t^v$, even though it alters the objective, does not alter the solution to \eqref{prob:InfGNEP} since it does not depend on the decision variables $(x^v,u^v)$.

The set of all possible solutions to \eqref{prob:InfGNEP} is characterized by the following \ac{gne} condition.
\begin{defn}[Generalized Nash equilibrium] \label{def:GeneralizedNashEquilibrium}\hfill\newline\null
    A \textit{\acl{gne}}, denoted $(\bx^*,\bu^*)$, is a solution to \eqref{prob:InfGNEP} where each agent $v$ minimizes its objective, given the other agents strategies $(\bx^{-v*},\bu^{-v*})$, such that
    \begin{equation*} \label{eq:GeneralizedNashEquilibrium}
        \!\sum\limits_{t=0}^{\infty} \bigl(\ell_t^v(x_t^{v\!}, \bx_t^{-v*\!}, u_t^{v\!}, \bu_t^{-v*\!}) - \underbrace{\ell_t^v(x_t^{v*\!}, \bx_t^{-v*\!}, u_t^{v*\!}, \bu_t^{-v*\!})\!\!}_{\smash{\ell_t^v(\bx_t^*, \bu_t^*)}}\;\bigr) \!\geq 0\!
    \end{equation*}
    holds for all  $t \in \N$ and $(x_t^v,u_t^v) \in \X^v(\bx_t^{-v*}) \times \U^v(\bu_t^{-v*})$ subject to the dynamics \eqref{eq:Dynamics}.
\end{defn}
For a more general definition of \acp{gnep} and \acp{gne} see, e.g., \cite{facchinei_generalized_2007}. By definition, the \ac{gne} is strategically stable, i.e., no agent involved can gain any advantage by unilaterally changing their strategy. Also note that a \ac{gne} is not necessarily unique.

The infinite-horizon Problem \eqref{prob:InfGNEP} may admit an unbounded optimal value of the objective. To mitigate this issue, we henceforth consider the shift $\xi_t^v =\ell_t^v(\bx_t^*,\bu_t^*)$ with respect to a time-varying \ac{gne} and introduce the shifted stage cost, following directly from Definition \ref{def:GeneralizedNashEquilibrium}
\begin{align} \label{eq:ShiftedStageCost}
    \sum\limits_{t=0}^{\infty} \underbrace{\ell_t^v(\bx_t,\bu_t) - \ell_t^v(\bx_t^*,\bu_t^*)}_{\displaystyle=: \widehat{\ell}_t^v(\bx_t,\bu_t)} = \sum\limits_{t=0}^{\infty} \widehat{\ell}_t^v(\bx_t,\bu_t) \geq 0.
\end{align}
This condition resolves the well-posedness issue by considering the sum of differences rather than the (potentially unbounded) objective itself. Instead of this shift one might as well work with the concept of overtaking optimality \parencite[cf. Sec.~1.2.2][]{blot_infinite-horizon_2014}.

Since Problem~\eqref{prob:InfGNEP} is generally not numerically tractable, we approximate it by repeatedly solving a truncated finite-horizon problem with horizon length $N \in \N$
\begin{equation} \label{prob:GNEP} \tag{$\mathscr{P}_N^v$}
    \!\forall v \in \V :\left\{\rule{0pt}{55pt}\right.
    \renewcommand\arraystretch{1.4}
    \begin{array}{@{\;}r@{\quad}l@{}}\displaystyle
    \min_{\!\!\!x_{\bullet|t}^v, u_{\bullet|t}^v\!\!\!} & \sum\limits_{k=0}^{N-1} \widehat{\ell}_{t+k}^v(\bx_{k|t},\bu_{k|t})\\
    \text{s.t.}   & \forall\, k \in \{0,\ldots,N-1\} \!\boldsymbol{:}\\
                  & x_{0|t}^v = x_t^v \in \X^v\bigl(\bx_t^{-v}\bigr),\\
                  & x_{k+1|t}^v = A_{t+k}^v x_{k|t}^v + B_{t+k}^v u_{k|t}^v,\\
                  & x_{k+1|t}^v \in \X^v\bigl(\bx_{k+1|t}^{-v}\bigr),\qquad\\
                  & u_{k|t}^v \in \U^v\bigl(\bu_{k|t}^{-v}\bigr).\qquad
    \end{array}\hspace{\linewidth minus 1fil}
\end{equation}
Applying the open-loop solution of \eqref{prob:GNEP} in receding-horizon fashion gives the implicit feedback law
\begin{equation*}
    \kappa^v(\bx_t) = u_{0|t}^{v*},\quad \rlap{$\forall v \in \V$,}
\end{equation*} and thus defines the \acl{rhg}.

Subsequently, we analyse the stability of this time-varying \ac{rhg}. More specifically, we investigate under which conditions the closed-loop trajectory obtained from applying $\kappa^v(\bx_t)$, converges to a \ac{gne} $(\bx^*,\bu^*)$ obtained from solving \eqref{prob:InfGNEP}. From here on we will also refer to $(\bx^*,\bu^*)$ as an infinite-horizon optimal trajectory.

\FloatBarrier
\section{Potential Games} \label{sec:PGs}

To this end, we temporarily assume that the set of agents is a singleton, restricting to a single agent, so that \eqref{prob:GNEP} reduces to a standard \ac{mpc} to which the stability results of \cite{grune_economic_2020} apply. Therein, they show $\mathbb{P}$-practical asymptotic stability for time-varying \acl{mpc} with respect to the corresponding infinite-horizon optimal trajectory $(x^*, u^*)$. The natural question that arises is how to extend these results to the game-theoretic setting.

An important class of games that can be posed as optimization problems rather than game-theoretic equilibrium seeking are \aclp{pg}. In \aclp{pg}, agents unknowingly and simultaneously minimize a higher level objective function $\boldsymbol{p}_N(\bx_t,\bu_t, t)$ which allows us to reformulate \eqref{prob:GNEP} by the following centralized equivalent \ac{ocp}
\begin{equation} \label{prob:CentralizedOCP} \tag{$\mathscr{P}_N$}
    \renewcommand\arraystretch{1.4}
    \begin{array}{@{}r@{\quad}l@{}}\displaystyle
    \boldsymbol{V}_{\!\!N}(\bx_t,t) := \min_{\!\!\!\!\bx_{\bullet|t},\bu_{\bullet|t}\!\!\!\!} & \boldsymbol{p}_N(\bx_{\bullet|t},\bu_{\bullet|t}, t)\\
    \text{s.t.}   & \forall\, k \in \{0,\ldots,N-1\} \!\boldsymbol{:}\\
                  & \bx_{0|t} = \bx_t \in \bX,\\
                  & \bx_{k+1|t} = \boldsymbol{A}_{t+k} \bx_{k|t} + \boldsymbol{B}_{t+k} \bu_{k|t},\\
                  & \bx_{k+1|t} \in \bX,\\
                  & \bu_{k|t} \in \bU.
    \end{array}\hspace{\linewidth minus 1fil}
\end{equation}
In order for Problem~\eqref{prob:GNEP} and \eqref{prob:CentralizedOCP} to be equivalent, the minimization of $\boldsymbol{p}_N(\bx,\bu, t)$ with respect to agent $v$'s decision variables must be equal to the minimization of agent $v$'s objective function. This requires that the gradient of $\boldsymbol{p}_N(\bx,\bu, t)$ with respect to each agent's decision variables coincides with the gradient of that agent's objective function. One can show that this only holds under a symmetric coupling between the agents objective functions \parencite[Theorem~4.5][]{monderer_potential_1996} which, imposed to our stage cost, yields the following assumption.
\begin{assum}[Symmetric off-diagonal terms] \label{assum:LQGame}\hfill\newline\null
    The off-diagonal terms of the block-sparse cost matrices are symmetric for all $t \in \N$ such that
    \begin{equation*}
        Q_t^{i,j} = \,\bigr(Q_t^{j,i}\bigr){}^{\!\top}, \quad R_t^{i,j} = \,\bigr(R_t^{j,i}\bigr){}^{\!\top}, \quad \forall\, i \neq j \in \V.
    \end{equation*}
\end{assum}
Given Assumption~\ref{assum:LQGame}, we can now explicitly construct $\boldsymbol{p}_N(\bx,\bu, t)$, also referred to as the \textit{potential function}.\!\footnote{See \cite{facchinei_decomposition_2011} for a formal treatment of generalized potential games.}
\begin{prop}
    Under Assumption~\ref{assum:LQGame} the potential function of Problem~\eqref{prob:GNEP}, with $N \in \N$, is given by
    \begin{equation}\label{eq:PotentialFunction}
        \boldsymbol{p}_N(\bx,\bu,t) := \smash{\sum_{k=0}^{N-1}} \boldsymbol{p}_{t+k}(\bx_{k|t},\bu_{k|t}),
    \end{equation}
    with
    \begin{equation*}
        \boldsymbol{p}_t(\bx_t,\bu_t) := \bx_t^\top \boldsymbol{Q}_t \bx_t + \bu_t^\top \boldsymbol{R}_t \bu_t + \boldsymbol{q}_t \bx_t + \boldsymbol{r}_t \bu_t,
    \end{equation*}
    where $\boldsymbol{Q}_t \!:=\! \frac{1}{2}\left(\sum_{v \in \V} \boldsymbol{Q}_t^v\right) + \frac{1}{2}\blkdiag\bigl(Q_t^{1,1}, \ldots, Q_t^{M,M}\bigr)$,\vspace{3pt}\linebreak
    $\boldsymbol{q}_t := \bigl[q_t^1,\ldots,q_t^M\bigr]$ and $\boldsymbol{R}_t$ and $\boldsymbol{r}_t$ defined analogously.
\end{prop}
\begin{pf}
    By \cite[Lemma~4.4][]{monderer_potential_1996} a continuously differentiable game is an exact potential game if
    \begin{equation*}
        \!\!\nabla_{\!\!\left(\rule{0pt}{6pt}\right.\! x_{\bullet|t}^v\!, u_{\bullet|t}^v\!\left.\rule{0pt}{6pt}\right)}\;\, \smash{\sum\limits_{k=0}^{N-1}} \bigl(\ell_{t+k}^v\bigl(\bx_{k|t},\bu_{k|t}\bigr) - \boldsymbol{p}_{t+k}\bigl(\bx_{k|t},\bu_{k|t}\bigr) \bigr) = 0
    \end{equation*}
    holds for all $v \in \V$ and all $t \in \N$. An equivalent condition is that, for all $v \in \V$ and all $t \in \N$, the following holds pointwise
    \begin{equation} \label{eq:Gradient}
        \!\!\nabla_{\!\!\left(\rule{0pt}{6pt}\right.\! x_t^v\!, u_t^v\!\left.\rule{0pt}{6pt}\right)}\;\, \bigl(\ell_t^v\bigl(\bx_t,\bu_t\bigr) - \boldsymbol{p}_t(\bx_t,\bu_t)\bigr) = 0.
    \end{equation}
    This is guaranteed under Assumption~\ref{assum:LQGame}, since
    \begin{equation*}
        \begin{alignedat}{1}
            \!\!\nabla_{x_t^v}\;\, \ell_t^v\bigl(\bx_t,\bu_t\bigr) =&\; \bigl(q_t^v\bigr){}^{\!\top} + \bigl(Q_t^{v,v} + \bigr(Q_t^{v,v}\bigr){}^{\!\top}\bigr) x_t^v
            \\&+ \smash{\sum_{\!i \,\in\, \V \setminus v\!\!\!}} Q_t^{v,i}\,x_t^i
        \end{alignedat}
    \end{equation*}
    and
    \begin{equation*}
        \begin{alignedat}{1}
            \!\!\nabla_{x_t^v}\;\, \boldsymbol{p}_t(\bx_t,\bu_t) =&\; \bigl(q_t^v\bigr){}^{\!\top} + \bigl(Q_t^{v,v} + \bigr(Q_t^{v,v}\bigr){}^{\!\top}\bigr) x_t^v
            \\&+ \frac{1}{2}\smash{\sum_{\!i \,\in\, \V \setminus v\!\!\!}} \bigl(Q_t^{v,i} + \bigr(Q_t^{i,v}\bigr){}^{\!\top}\bigr) \,x_t^i
        \end{alignedat}
    \end{equation*}
    are identical which similarly holds true for the gradients with respect to the input $u_t^v$.
\end{pf}

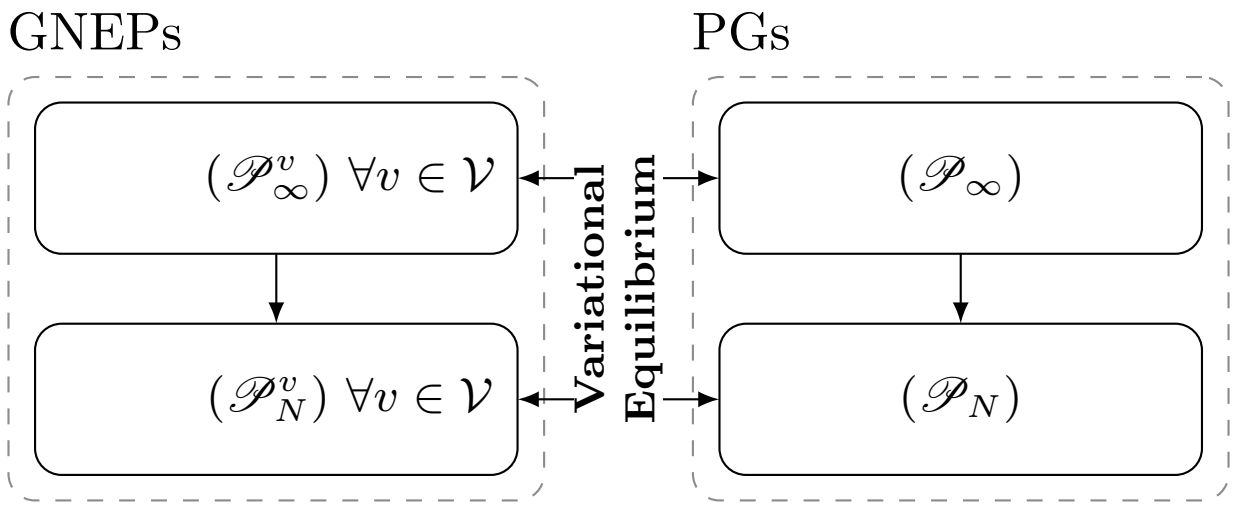
\begin{figure}[ht]
    \centering
    \input{figures/problems}
    \caption{Connection between the \acp{gnep} and the centralized potential games for the infinite-horizon as well as the receding-horizon case.}
    \label{fig:Problems}
\end{figure}

The following corollary follows immediately, since \eqref{eq:Gradient} holds pointwise for all $t$ and therefore also for the infinite-horizon case $N \rightarrow \infty$.
\begin{cor}
    Under Assumption~\ref{assum:LQGame} the potential function of Problem~\eqref{prob:InfGNEP} is given by \eqref{eq:PotentialFunction} with $N \rightarrow \infty$.
\end{cor}
Accordingly, we can reformulate \eqref{prob:InfGNEP} as follows
\begin{equation} \label{prob:InfCentralizedOCP} \tag{$\mathscr{P}_{\infty}$}
    \renewcommand\arraystretch{1.4}
    \begin{array}{@{}r@{\quad}l@{}}\displaystyle
    \widehat{\boldsymbol{V}}_{\!\!\infty}(\bx_t,t) := \min_{\!\!\!\!\bx,\bu\!\!\!\!} & \widehat{\boldsymbol{p}}_{\infty}(\bx,\bu, t)\\
    \text{s.t.}   & \forall\, t \in \N \!\boldsymbol{:}\\
                  & \bx_{t+1} = \boldsymbol{A}_t \bx_t + \boldsymbol{B}_t \bu_t,\\
                  & \bx_t \in \bX,\\
                  & \bu_t \in \bU,
    \end{array}\hspace{\linewidth minus 1fil}
\end{equation}
with $\widehat{\boldsymbol{p}}_{\infty}(\bx_{\bullet|t},\bu_{\bullet|t}, t)$ being the shifted potential function where
\begin{equation*}\label{eq:ShiftedStagePotentialFunction}
    \widehat{\boldsymbol{p}}_t(\bx_t,\bu_t) := \boldsymbol{p}_t(\bx_t,\bu_t) - \boldsymbol{p}_t(\bx_t^*,\bu_t^*)
\end{equation*}
denotes the respective shifted stage potential function analogously to \eqref{eq:ShiftedStageCost}.
Figure \ref{fig:Problems} summarizes the different problems introduced and how they relate to each other.

Generally, the equivalence between \eqref{prob:InfGNEP} and \eqref{prob:InfCentralizedOCP} (\eqref{prob:GNEP} and \eqref{prob:CentralizedOCP}, respectively) does not necessarily hold true, in the sense that not all \acp{gne} of \eqref{prob:InfGNEP} are naturally preserved in \eqref{prob:InfCentralizedOCP} (\eqref{prob:CentralizedOCP}, respectively) \parencites[see][]{facchinei_decomposition_2011}[Section~3.2][]{le_cadre_parametrized_2022}. In fact, even in the case of strict convexity, where \eqref{prob:CentralizedOCP} admits a unique solution, the set of \acp{gne} of \eqref{prob:GNEP} does not necessarily reduce to a singleton (see, e.g., \cite[Example~1][]{facchinei_generalized_2007}). This one-to-one correspondence is only given for strictly convex games without coupling constraints i.e. (regular) \acp{nep} \parencites[Corollary~1][]{facchinei_generalized_2007}[Corollary~1][]{neyman_correlated_1997}.

This is why we introduce a refined solution concept, also known as the variational equilibrium which is indeed nothing else than a stationary point to \eqref{prob:InfCentralizedOCP} and \eqref{prob:CentralizedOCP} \parencite[Chp.~1.3.1][]{facchinei_finite-dimensional_2003}.
\begin{defn}[Variational equilibrium] \label{def:VGNE}\hfill\newline\null
    Let the stage cost \eqref{eq:StageCost} be strictly convex for all $v \in \V$ and let $N \in \N \cup \{\infty\}$. A trajectory pair $(\bx^*,\bu^*)$ solving \eqref{prob:InfCentralizedOCP}, \eqref{prob:CentralizedOCP}, respectively, is a \textit{\ac{vgne}} satisfying
    \begin{equation*} \label{eq:VariationalInequality} \tag{VI}
        \begin{bmatrix}
            \bx-\bx^*\\
            \bu-\bu^*
        \end{bmatrix}^\top\!\!
        \nabla_{\left(\bx^*,\bu^*\right)}\; \boldsymbol{p}_{N}(\bx^*,\bu^*,t) \geq 0,
    \end{equation*}
    for all  $t \in \N$ and $(\bx_t,\bu_t) \in \bX \times \bU$ subject to the dynamics \eqref{eq:Dynamics}.
\end{defn}
In other words, given strict convexity, the solution of \eqref{prob:InfCentralizedOCP} and \eqref{prob:CentralizedOCP}, is equivalent to the solution of the variational inequality \eqref{eq:VariationalInequality}. Note that every \ac{vgne} is also a \ac{gne} but not vice-versa.

The convexity requirement of Definition~\ref{def:VGNE}, together with the uniqueness needed for a well-posed \eqref{eq:VariationalInequality} and a meaningful stability analysis is formalized in the following assumption.
\begin{assum}[Uniqueness and strict convexity] \label{assum:UniquenessAndStrictConvexity}
    We assume that the cost matrices $(\boldsymbol{Q}_t$, $\boldsymbol{R}_t)$ are uniformly positive definite, i.e., $\boldsymbol{Q}_t, \boldsymbol{R}_t \succeq \underline{c}\,I$ for some $\underline{c} > 0$ and all $t \in \N$, such that $\boldsymbol{p}_t(\bx_t,\bu_t)$ is strictly convex, and that the solution to \eqref{prob:InfCentralizedOCP} is unique, such that there exists only one infinite-horizon optimal trajectory $(\bx^*,\bu^*)$.
\end{assum}

Henceforth, using the \ac{pg}-reformulation, we are restricting ourselves to the study of \acp{vgne} and decomposition algorithms for \eqref{prob:GNEP} that also converge to a \ac{vgne} rather than a more general \ac{gne} \parencite[see, e.g.][]{ananduta_bregman_2025,migot_decomposition_2021}. An interesting discussion of the implications of the \ac{vgne} and its fairness, together with possible improvements can be found in \cite{hall_limits_2025}.

Having established the centralized reformulation, we can now proceed to analyse the stability of \eqref{prob:GNEP} based on \eqref{prob:CentralizedOCP}.

\section{Verification of Closed-Loop Stability Results} \label{sec:Stability}

In order to derive $\mathbb{P}$-practical asymptotic stability we are building on the results of \cite{grune_economic_2020} providing a new set of sufficient conditions leveraging our \acl{ltv} dynamics, quadratic objective and convex constraints. Here, two central concepts utilised are strict dissipativity, and the turnpike property which we define according to \cite[Proposition~4.1][]{faulwasser_economic_2018}.

\begin{figure}[b]
    \centering
    \includegraphics[width=230pt]{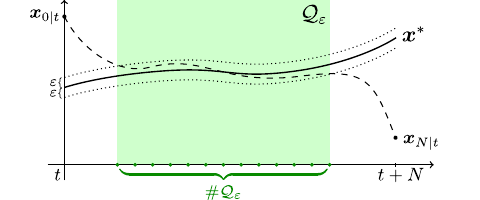}
    \caption{Time-Varying turnpike behaviour of $\bx_{\bullet|t}$ w.r.t $\bx^*$ (Figure adapted from \cite[Figure~4.2][]{pirkelmann_economic_2020}).}
    \label{fig:turnpike}
\end{figure}

\begin{defn}[Turnpike property] \label{def:Turnpike}
    Problem~\eqref{prob:CentralizedOCP} exhibits the turnpike property if there exists $C < \infty$ and $\alpha(\varepsilon) \in \mathcal{K}_\infty$ such that
    \begin{equation*}
        \# \mathcal{Q}_\varepsilon \geq N - \frac{C}{\alpha(\varepsilon)}
    \end{equation*}
    with $\mathcal{Q}_{\varepsilon\!} \!:=\! \!\{k \!\in\! \{0,\ldots,\!N\shortminus1\;\!\!\} \!\mid\! (\bx_{k|t},\! \bu_{k|t}) \!\in\! \mathcal{B}_{\varepsilon}(\bx_{t+k}^*,\! \bu_{t+k}^*)\;\!\!\}\!$ the set of all times $k$ where the open-loop trajectory lies within an $\varepsilon$-vicinity of the infinite-horizon optimal trajectory $(\bx^*, \bu^*)$. $\#\mathcal{Q}_\varepsilon$ denotes the cardinality of $\mathcal{Q}_\varepsilon$, which corresponds to the total time spent \textit{within this $\varepsilon$-vicinity} around the optimal trajectory.
\end{defn}
Figure \ref{fig:turnpike} depicts this behaviour for a general time-varying infinite-horizon optimal trajectory $\bx^*$ and an open-loop prediction $\bx_{\bullet|t}$ starting off the optimal trajectory.

Strict dissipativity is closely related to the turnpike property, and in fact, under the right set of assumptions, it can be shown, that it implies turnpike behaviour \parencite{grune_strict_2018}. Similar to \cite[Definition~7][]{grune_strict_2018} we introduce time-varying strict dissipativity for the centralized Problem~\eqref{prob:CentralizedOCP}.
\begin{defn}[Time-varying strict dissipativity] \label{def:StrictDissipativity}
    Problem~\eqref{prob:CentralizedOCP} is time-varying strictly dissipative with respect to an infinite-horizon optimal trajectory $(\bx^*,\bu^*)$ and the supply rate $\widehat{\boldsymbol{p}}_t(\bx_t, \bu_t)$ if there exists a time-varying storage function $\boldsymbol{\Lambda}_t(\bx_t) : \N \times \domain{n} \mapsto \domain{}$ bounded from below, such that\vspace{-5pt}
    \begin{multline} \label{eq:StrictDissipativity}
        \boldsymbol{\Lambda}_{t+1}(\boldsymbol{A}_t \bx_t + \boldsymbol{B}_t \bu_t) - \boldsymbol{\Lambda}_t(\bx_t) \hspace{\linewidth minus 1fil}\\
        \leq -\boldsymbol{\alpha}(\|(\bx_t-\bx_t^*, \bu_t-\bu_t^*)\|) + \widehat{\boldsymbol{p}}_t(\bx_t,\bu_t)
    \end{multline}
    holds for all $t \in \N$, with $\boldsymbol{\alpha}(\|(\bx_t-\bx_t^*, \bu_t-\bu_t^*)\|) \in \mathcal{K}_\infty$.
    For the case of $\boldsymbol{\alpha}(\|(\bx_t-\bx_t^*, \bu_t-\bu_t^*)\|) = 0$ the above notation is also referred to as time-varying dissipativity.
\end{defn}
For different dissipativity notions, cf. \cite{muller_dissipativity_2021,kohler_periodic_2018}.

We further introduce the rotated problem as a tool for stability analysis \parencite{grune_economic_2020}. The rotated stage potential function is given by\vspace{-5pt}
\begin{equation} \label{eq:RotatedStagePotentialFunction}
    \!\!\!\lowtilde{\boldsymbol{p}}_t(\bx_t,\!\bu_t) \,\!\!:= \underbrace{\widehat{\boldsymbol{p}}_t(\bx_t,\!\bu_t) \!+\! \boldsymbol{\Lambda}_t(\bx_t) \!-\! \boldsymbol{\Lambda}_{t+1}(\boldsymbol{A}_t \bx_t \!+\! \boldsymbol{B}_t \bu_t)\!}_{\stackrel{\smash{\eqref{eq:StrictDissipativity}}}{\geq}\,\boldsymbol{\alpha}(\|(\bx_t-\bx_t^*, \bu_t-\bu_t^*)\|)}\!\!\vspace{-5pt}
\end{equation}
with the corresponding rotated potential function\vspace{-10pt}
\begin{align*}
    \lowtilde{\boldsymbol{p}}_N(\bx,\bu,t) &:= \!\sum_{k=t}^{\!\!t+N-1\!\!} \lowtilde{\boldsymbol{p}}_k(\bx_k,\bu_k)\\
                                        &\phantom{:}= \!\sum_{k=t}^{\!\!t+N-1\!\!} \widehat{\boldsymbol{p}}_k(\bx_k,\bu_k) + \boldsymbol{\Lambda}_t(\bx_t) \!-\! \boldsymbol{\Lambda}_{t+N}(\bx_{t+N})\notag\\[-25pt]\notag
\end{align*}
and the rotated optimal value function denoted by $\lowtilde{\boldsymbol{V}}_{\!\!N}(\bx_t,t)$, obtained from solving Problem~\eqref{prob:CentralizedOCP} modified with $\lowtilde{\boldsymbol{p}}_N(\bx,\bu,t)$ as the objective function.\!\footnote{In the literature the rotated problem is sometimes also referred to as the modified problem.} Henceforth we will denote the rotated problem by $(\lowtilde{\mathscr{P}}_N)$.

We proceed as outlined in Figure~\ref{fig:Assumptions} by first establishing time-varying strict dissipativity and continuity of the respective storage function using an \textit{optimality condition} and a \textit{Slater-like assumption}. From there on we show turnpike behaviour for both the original Problem~\eqref{prob:CentralizedOCP} and its rotated counterpart $(\lowtilde{\mathscr{P}}_N)$ using a \textit{reachability assumption}. Following that, we show continuity of the stage potential function $\boldsymbol{p}_t(\bx_t,\bu_t)$ exploiting our problem structure, from which we also deduce an upper bound on the rotated stage potential function $\lowtilde{\boldsymbol{p}}_t(\bx_t,\bu_t)$. Together with a \textit{controllability argument} we finally derive (approximate) continuity of the optimal value function for both the original and the rotated problem which allows us to invoke the stability results of \cite{grune_economic_2020}.

We begin by establishing time-varying strict dissipativity. To this end, we rely on a connection between strong duality and strict dissipativity that was first observed by \cite{diehl_lyapunov_2011,angeli_average_2012}, formalized by \cite{damm_exponential_2014} and extended to the time-varying but unconstrained setting by \cite{grune_strict_2018}. We further extend these results to include convex, time-invariant constraints building on a strong Pontryagin principle given by \cite{blot_infinite-horizon_2007}. Therefore, we state the following optimality condition.

\begin{assum}[Optimality condition] \label{assum:OptimalityCondition} We assume
    \begin{enumerate}[label=\textnormal{(\roman*)}]
        \item the state transition matrix $A_t^v$ has full rank for all $v \in \V$ and $t \in \N$.
        \item the state inequality constraints $\boldsymbol{g}_{\bx}(\bx_t)$ are affine, and $\boldsymbol{g}_{\bu}(\bu_t)$ is at least twice differentiable.\label{assum:Lipschitz}
        \item \eqref{prob:InfCentralizedOCP} is non-degenerate, meaning, its objective multiplier $\eta_0$ is non-zero.
        \item the state  and constraint multipliers of \eqref{prob:InfCentralizedOCP} are uniformly bounded by $\overline{\lambda} \!:=\! \sup_{t \in \N} \|\lambda_t\| \!<\! \infty$ and $\overline{\vartheta} \!:=\! \sup_{t \in \N} \|\lowtilde{\vartheta}_t\| \!<\! \infty$, respectively.\label{assum:BoundedMultipliers}
    \end{enumerate}
\end{assum}
The next Slater-like assumption is needed to guarantee pointwise bounded multipliers in Lemma~\ref{lem:StrongMaximumPrinciple} \parencite[see][]{blot_infinite-horizon_2007}.
\begin{assum}[Slater-like condition]\label{assum:SlaterlikeCondition}\hfill\newline\null
    For all $t \in \N$, there exists a strictly feasible point $(\accentset{\frown}{\bx}_t, \accentset{\frown}{\bu}_t)$ such that
    \begin{equation*}
        \emptyset \neq \left\{(\accentset{\frown}{\bx}_t,\accentset{\frown}{\bu}_t) \mid \boldsymbol{g}(\accentset{\frown}{\bx}_t,\accentset{\frown}{\bu}_t) < 0, \boldsymbol{A}_t \accentset{\frown}{\bx}_t + \boldsymbol{B}_t \accentset{\frown}{\bu}_t = \bx_{t+1}^*\right\}
    \end{equation*}
    holds.
\end{assum}
This assumption is satisfied, if around the optimal trajectory $(\bx^*,\bu^*)$, there exists a one-step reachable set on which no inequality constraint is active.

\begin{rem}
    Interestingly, constraint qualifications normally utilised for finite-horizon optimality conditions (such as, e.g., KKT or Fritz John conditions \parencite{bertsekas_nonlinear_1999}) like LICQ \parencite{wachsmuth_licq_2013} or MFCQ \parencite{gauvin_necessary_1977} do not naturally carry over to the infinite-horizon setting, as the underlying rank and compactness arguments do not extend to the infinite-dimensional multiplier spaces involved. Instead, one must resort to different regularity conditions such as Assumption~\ref{assum:SlaterlikeCondition}. Alternatively, \cite[Theorem~2.19, Condition~(3)][]{blot_infinite-horizon_2014} provides an LICQ-like condition, which turns out, however, to be much more restrictive. Moreover, as shown by \cite{halkin_necessary_1974} in continuous time, the usual transversality condition on the costate does not necessarily hold on infinite-horizons. This problem also persists in infinite-horizon discrete-time problems \parencite{aseev_optimality_2016}, which is precisely why we rely on Assumption~\ref{assum:OptimalityCondition}~Condition~\ref{assum:BoundedMultipliers}.
\end{rem}

Under Assumption~\ref{assum:OptimalityCondition} and \ref{assum:SlaterlikeCondition}, we can now restate \cite[Theorem~3][]{blot_infinite-horizon_2007}, adapted to our notation. Following \cite{blot_infinite-horizon_2007} we write the dynamics compactly as $\boldsymbol{f}_t(\bx_t,\bu_t) := \boldsymbol{A}_t\bx_t + \boldsymbol{B}_t\bu_t$ and reformulate the constraints as
\begin{equation*}
    \lowtilde{\boldsymbol{g}}(\bx_{t+1}, \bu_t) := 
    \begin{bmatrix}
        \boldsymbol{g}_{\bx}\left(\smash{\overbrace{\boldsymbol{A}_t^{-1}(\bx_{t+1} - \boldsymbol{B}_t\bu_t)}^{\bx_t}}\right) \\
        \boldsymbol{g}_{\bu}(\bu_t)
    \end{bmatrix} \leq 0,
\end{equation*}
which lets us restate their result in form of the following preparatory lemma, which formulates an infinite-horizon discrete-time minimum principle.
\begin{lem}[\cite{blot_infinite-horizon_2007}] \label{lem:StrongMaximumPrinciple}
    Let $(\bx^*, \bu^*)$ be an infinite-horizon optimal trajectory of \eqref{prob:InfCentralizedOCP} and let the following conditions hold.
    \begin{enumerate}[label=\textnormal{(\alph*)}]
        \item \label{cond:InvertibleStateTransitionMatrix}
        For all $t \in \N$, the state transition matrix $\boldsymbol{A}_t$ is invertible.\vspace{2pt}
        \item \label{cond:Differentiability}
        For all $t \in \N$, $\frac{\partial \boldsymbol{p}_t}{\partial \bx}\Bigl|_{\raisebox{2pt}{$\scriptstyle(\bx_t^*, \bu_t^*)$}}$ and $\frac{\partial \lowtilde{\boldsymbol{g}}}{\partial \bx}\Bigl|_{\raisebox{2pt}{$\scriptstyle(\bx_{t+1}^*, \bu_t^*)$}}$ exists.
        \item \label{cond:IoffeTihomirov}
        For all $t \in \N$, the Ioffe-Tihomirov\footnote{This condition is an equivalent condition for Michel's convexity condition \parencite{michel_programmes_1980} and is automatically satisfied under convex objective, convex constraints, and \ac{ltv} dynamics \parencite{blot_infinite-horizon_2007}.\label{foot:IoffeTihomirov}}\! condition holds.
        \item \label{cond:PointwiseSlater}
        For all $t \in \N$, there exists a $\lowtilde{\bu}_t \in \bU$ such that\hfill\newline\null$\lowtilde{\boldsymbol{g}}(\bx_{t+1}^*, \lowtilde{\bu}_t) < 0$.\vspace{-5pt}
    \end{enumerate}
    Then, there exist multipliers $\eta_0 \in \domain{}$, $\lambda_{t+1} \in \domain{n}$ and $\lowtilde{\mu}_t \in \domain{c}$ satisfying the following conditions\vspace{-5pt}
    \begin{enumerate}[label=\textnormal{(\roman*)}]
        \item \label{cond:WellPosedness}$\left(\eta_0, \lambda_1\right) \neq(0,0)$.
        \item \label{cond:GreaterZero}$\,\eta_0 \geq 0$, $\lowtilde{\mu}_t \geq 0$ and $\lowtilde{\mu}_t^\top \lowtilde{\boldsymbol{g}}(\bx_{t+1}^*, \bu_t^*) = 0$ for all $t \in \N$.
        \item For all $t \in \N$,\label{cond:OptimalityInX}
            \begin{equation*}
                \lambda_t^\top \!\!=
                \lambda_{t+1}^\top \boldsymbol{A}_t+\,
                \eta_0 \smash{\frac{\partial \boldsymbol{p}_t}{\partial \bx}\biggl|_{(\bx_t^*, \bu_t^*)}}       \!\!\!\!\!\!+\, \lowtilde{\mu}_{t\shortminus1}^\top \smash{\frac{\partial \lowtilde{\boldsymbol{g}}}{\partial \bx}\biggl|_{(\bx_t^*, \bu_{t\shortminus1}^*)}}.\vspace{5pt}
            \end{equation*}
        \item For all $t \in \N$, for all $\bu_t \in \bU$,\vspace{-6pt} \label{cond:OptimalityInU}
            \begin{align*}
                &\eta_0\, \boldsymbol{p}_{t}(\bx_t^*,\bu_t^*) + \lambda_{t+1}^\top \boldsymbol{f}_t(\bx_t^*,\bu_t^*) + \lowtilde{\mu}_t^\top \lowtilde{\boldsymbol{g}}(\bx_{t+1}^*,\bu_t^*)\\
                &\leq\eta_0\, \boldsymbol{p}_{t}(\bx_t^*,\bu_t) + \lambda_{t+1}^\top \boldsymbol{f}_t(\bx_t^*,\bu_t) + \lowtilde{\mu}_t^\top \lowtilde{\boldsymbol{g}}(\bx_{t+1}^*,\bu_t).
            \end{align*}
    \end{enumerate}
\end{lem}
It can easily be seen that Conditions~\ref{cond:InvertibleStateTransitionMatrix}--\ref{cond:IoffeTihomirov} are readily satisfied under Assumption~\ref{assum:OptimalityCondition}, while Condition~\ref{cond:PointwiseSlater} requires that for all $t \in \N$, the following holds
\begin{equation*}
    \emptyset \neq \left\{\lowtilde{\bu}_t \mid \lowtilde{\boldsymbol{g}}(\bx_{t+1}^*,\lowtilde{\bu}_t) < 0\right\}
\end{equation*}
which is exactly our Slater-like condition (Assumption~\ref{assum:SlaterlikeCondition}) and hence, in our setting, all conditions of Lemma~\ref{lem:StrongMaximumPrinciple} are satisfied.

 Note that Condition~\ref{cond:OptimalityInU} can equivalently be expressed by the following first-order necessary condition
\begin{equation}\label{eq:FirstOrderCondition}
     0 = \eta_0 \smash{\frac{\partial \boldsymbol{p}_t}{\partial \bu}\biggl|_{(\bx_t^*, \bu_t^*)}} \!\!\!\!\!\!+\,
     \lambda_{t+1}^\top \boldsymbol{B}_t+\, \lowtilde{\mu}_t^\top \smash{\frac{\partial \lowtilde{\boldsymbol{g}}}{\partial \bu}\biggl|_{(\bx_{t+1}^*, \bu_t^*)}}\!\!\!\!
\end{equation}
which will turn out to be useful later. We now come to our main result.

\begin{thm} \label{thm:StrictDissipativity}
    Let Assumption~\ref{assum:UniquenessAndStrictConvexity}, \ref{assum:OptimalityCondition} and \ref{assum:SlaterlikeCondition} hold and suppose $N \in \N$. Then Problem~\eqref{prob:CentralizedOCP} is time-varying strictly dissipative with respect to the shifted stage potential function $\widehat{\boldsymbol{p}}_t(\bx_t,\bu_t)$ and the infinite-horizon optimal trajectory $(\bx^*, \bu^*)$ obtained from solving \eqref{prob:InfCentralizedOCP}.
\end{thm}

\begin{pf}
    In order to show time-varying strict dissipativity with respect to the shifted stage potential function $\widehat{\boldsymbol{p}}_t(\bx_t,\bu_t)$ we derive a storage function $\boldsymbol{\Lambda}_t(\bx_t)$ such that the dissipation inequality \eqref{eq:StrictDissipativity} holds for all $t \in \N$. To this end, we propose the following time-varying candidate storage function
    \begin{equation} \label{eq:StorageFunction}
        \!\!\!\boldsymbol{\Lambda}_t(\bx_t) \!:=\! \smash{\dfrac{\lowtilde{\vartheta}_{t\shortminus1\!}^\top}{\eta_0}}\boldsymbol{g}_{\bx\!}\left(\boldsymbol{A}_{t\shortminus1\!}^{-1}(\bx_t \!-\! \boldsymbol{B}_{t\shortminus1\!}\bu_{t\shortminus1}^*)\right) - \smash{\dfrac{\lambda_t^\top\!\!}{\eta_0\!\!}}(\bx_t - \bx_t^*)\!\!\!
    \end{equation}
    where $\lowtilde{\mu}_t =: \bigl(\lowtilde{\vartheta}_t,\lowtilde{\zeta}_t\bigr)$, with $\eta_0$, $\lambda_t^\top$ and $\lowtilde{\vartheta}_t$ the cost, state and state-inequality constraint multipliers from Lemma~\ref{lem:StrongMaximumPrinciple}. Proving the dissipation inequality \eqref{eq:StrictDissipativity} therefore reduces to establishing the lower bound on the rotated stage potential function $\lowtilde{\boldsymbol{p}}_t(\bx_t,\bu_t)$ (see \eqref{eq:RotatedStagePotentialFunction}). To this end, we substitute the above ansatz into \eqref{eq:RotatedStagePotentialFunction}, yielding
    \begin{equation*} \label{eq:ScaledRotatedStageCost}
        \begin{alignedat}{2}
            \!\!\smash{\underbrace{\eta_0\,\lowtilde{\boldsymbol{p}}_t(\bx_t,\bu_t)}_{  =:\overline{\boldsymbol{p}}_t(\bx_t,\bu_t)}} &= \eta_0\,\widehat{\boldsymbol{p}}_t(\bx_t,\bu_t) - \lambda_t^\top\!(\bx_t - \bx_t^*)\\
            &\quad\!+\lowtilde{\vartheta}_{t-1}^\top\boldsymbol{g}_{\bx}\!\left(\boldsymbol{A}_{t-1}^{-1}(\bx_t \!-\! \boldsymbol{B}_{t-1}\bu_{t-1}^*)\right)\!\!\\
            &\qquad\!\!+ \lambda_{t+1}^\top(\boldsymbol{A}_t\bx_t + \boldsymbol{B}_t\bu_t - \bx_{t+1}^*)\\
            &\qquad\quad\!\!\!-\lowtilde{\vartheta}_t^\top\!\boldsymbol{g}_{\bx}\!\left(\bx_t \!+\! \boldsymbol{A}_t^{-1\!}\boldsymbol{B}_t(\bu_t \!-\! \bu_t^*)\right)\!\!\!
        \end{alignedat}
    \end{equation*}
    where $\overline{\boldsymbol{p}}_t(\bx_t,\bu_t)$ denotes the scaled, rotated stage potential function. Since we assume $\eta_0 \neq 0$, and from Lemma~\ref{lem:StrongMaximumPrinciple}~Condition \ref{cond:GreaterZero} we conclude that $\eta_0 > 0$ such that this scaling is well-defined.

    Subsequently, we show that $\lowtilde{\boldsymbol{p}}_t(\bx_t,\bu_t)$ can be lower-bounded by $\boldsymbol{\alpha}(\|(\bx_t-\bx_t^*, \bu_t-\bu_t^*)\|)$, which we obtain via an intermediate bound on $\overline{\boldsymbol{p}}_t(\bx_t,\bu_t)$. First, we demonstrate pointwise in time that $\overline{\boldsymbol{p}}_t(\bx_t,\bu_t)$ attains its unique minimum at the infinite-horizon optimal trajectory $(\bx^*, \bu^*)$. Then we establish a lower bound $\boldsymbol{\alpha}(\|(\bx_t-\bx_t^*, \bu_t-\bu_t^*)\|)$.

    The minimum of $\overline{\boldsymbol{p}}_t(\bx_t,\bu_t)$ is obtained via the following constraint \ac{ocp}, which can be equivalently expressed through its dual problem given that strong duality holds
    \begin{equation} \label{eq:KKTOCP}
        \!\!\vcenter{\hbox{\input{figures/kkt}}}\!\!
    \end{equation}
    where $\boldsymbol{L}_t(\bx_t,\bu_t,\mu_t)$ denotes the Lagrangian
    \begin{equation*}
        \boldsymbol{L}_t(\bx_t,\bu_t,\mu_t) := \overline{\boldsymbol{p}}_t(\bx_t,\bu_t) + \mu_t^\top \boldsymbol{g}(\bx_t,\bu_t),
    \end{equation*}
    and $\mu_t \in \domain{c}$ denotes its multiplier associated with the joint state and input constraints. By Assumption~\ref{assum:UniquenessAndStrictConvexity}, $\boldsymbol{p}_t(\bx_t,\bu_t)$ is strictly convex, and hence so is $\overline{\boldsymbol{p}}_t(\bx_t,\bu_t)$, since it is the sum of strictly convex and affine functions \parencite[Lemma~2.1.6][]{nesterov_lectures_2018}. Together with the convexity of the constraints $\bX \times \bU$, Problem~\eqref{eq:KKTOCP} is strictly convex. Assumption~\ref{assum:SlaterlikeCondition} guarantees a strictly feasible point, so the \ac{kkt} conditions are necessary and sufficient for primal-dual optimality of $(\bx^*,\bu^*)$ \parencite[Chp.~5.5.3][]{boyd_convex_2004}.

    We first check the stationarity condition by deriving the gradient of the Lagrangian $\boldsymbol{L}_t(\bx_t,\bu_t,\mu_t)$ w.r.t. state
    \begin{subequations}\label{eq:StationarityI}
    \begin{flalign}
        \!\partderiv{\boldsymbol{L}_t}{\bx}\biggl|_{(\bx_t, \bu_t)} \hspace{-10pt}%
        &=\eta_0 \partderiv{\widehat{\boldsymbol{p}}_t}{\bx}\biggl|_{(\bx_t, \bu_t)} \hspace{-30pt}&&-\, \lambda_t + \lambda_{t+1}^\top \boldsymbol{A}_t + \lowtilde{\vartheta}_{t\shortminus1}^\top\partderiv{\boldsymbol{g}_{\bx}}{\bx}\biggl|_{\bx_{t\shortminus1}}\hspace{-9pt}\boldsymbol{A}_{t\shortminus1}^{-1} \!\!\!\!\!\!&\notag\\
        &&&- \lowtilde{\vartheta}_t^\top \partderiv{\boldsymbol{g}_{\bx}}{\bx}\biggl|_{\bx_t} +\; \mu_t^\top \partderiv{\boldsymbol{g}}{\bx}\biggl|_{\rlap{$\scriptstyle(\bx_t, \bu_t)$}}&\label{eq:Lx}
    \end{flalign}
    and input
    \begin{flalign}
        \!\partderiv{\boldsymbol{L}_t}{\bu}\biggl|_{(\bx_t, \bu_t)} \hspace{-10pt}
        &=\eta_0 \partderiv{\widehat{\boldsymbol{p}}_t}{\bu}\biggl|_{(\bx_t, \bu_t)} \hspace{-30pt}&&+\; \lambda_{t+1}^\top \boldsymbol{B}_t - \lowtilde{\vartheta}_t^\top\partderiv{\boldsymbol{g}_{\bx}}{\bx}\biggl|_{\bx_t}\hspace{-5pt}\boldsymbol{A}_t^{-1}\boldsymbol{B}_t &\notag\\
        & &&+ \mu_t^\top \partderiv{\boldsymbol{g}}{\bu}\biggl|_{(\bx_t, \bu_t)}.&\label{eq:Lu}
    \end{flalign}
    \end{subequations}
    Next, we note that the derivative of $\lowtilde{\mu}_t^\top\lowtilde{\boldsymbol{g}}(\bx_{t+1},\bu_t)$ can be expressed as
    \begin{align*}
        \lowtilde{\mu}_{t\shortminus1}^\top\partderiv{\lowtilde{\boldsymbol{g}}}{\bx}\biggl|_{(\bx_t, \bu_{t\shortminus1})} \hspace{-10pt}&= \;\lowtilde{\vartheta}_{t\shortminus1}^\top\partderiv{\boldsymbol{g}_{\bx}}{\bx}\biggl|_{\rlap{$\scriptstyle\bx_{t\shortminus1}$}}\;\, \boldsymbol{A}_{t\shortminus1}^{-1},\\
        \lowtilde{\mu}_t^\top\partderiv{\lowtilde{\boldsymbol{g}}}{\bu}\biggl|_{(\bx_{t+1}, \bu_t)} \hspace{-12pt}&= -\lowtilde{\vartheta}_t^\top\partderiv{\boldsymbol{g}_{\bx}}{\bx}\biggl|_{\rlap{$\scriptstyle\bx_t$}}\;\, \boldsymbol{A}_t^{-1}\boldsymbol{B}_t + \lowtilde{\zeta}_t^\top \partderiv{\boldsymbol{g}_{\bu}}{\bu}\biggl|_{\bu_t}
    \end{align*}
    and that $\lowtilde{\mu}_t\geq0$ (from Lemma~\ref{lem:StrongMaximumPrinciple}) is an admissible multiplier for \eqref{eq:KKTOCP} since $\lowtilde{\mu}_t^\top \lowtilde{\boldsymbol{g}}(\bx_{t+1}^*, \bu_t^*) = \lowtilde{\mu}_t^\top \boldsymbol{g}(\bx_t^*, \bu_t^*) = 0$, which allows us to rewrite \eqref{eq:StationarityI} as
    \begin{subequations}\label{eq:StationarityII}
    \begin{flalign}
        \!\partderiv{\boldsymbol{L}_t}{\bx}\biggl|_{(\bx_t, \bu_t)} \hspace{-10pt}%
        &=\eta_0 \partderiv{\widehat{\boldsymbol{p}}_t}{\bx}\biggl|_{(\bx_t, \bu_t)} \hspace{-13pt}-\, \lambda_t + \lambda_{t+1}^\top \boldsymbol{A}_t + \lowtilde{\mu}_{t\shortminus1}^\top \smash{\frac{\partial \lowtilde{\boldsymbol{g}}}{\partial \bx}\biggl|_{\rlap{$\scriptstyle(\bx_t, \bu_{t\shortminus1}\!)$}}} \hspace{-10pt}&\\
        \!\partderiv{\boldsymbol{L}_t}{\bu}\biggl|_{(\bx_t, \bu_t)} \hspace{-10pt}
        &=\eta_0 \partderiv{\widehat{\boldsymbol{p}}_t}{\bu}\biggl|_{(\bx_t, \bu_t)} \hspace{-13pt}+\; \lambda_{t+1}^\top \boldsymbol{B}_t + \lowtilde{\mu}_t^\top \partderiv{\lowtilde{\boldsymbol{g}}}{\bu}\biggl|_{\rlap{$\scriptstyle(\bx_{t+1}, \bu_t)$}}\;.&
    \end{flalign}
    \end{subequations}

    Evaluating \eqref{eq:StationarityII} at the infinite-horizon optimal trajectory $(\bx^*,\bu^*)$ and invoking Condition~\ref{cond:OptimalityInX} of Lemma~\ref{lem:StrongMaximumPrinciple} yields the state stationarity condition. The input stationarity condition follows analogously from Condition~\ref{cond:OptimalityInU} and \eqref{eq:FirstOrderCondition}
    \begin{equation} \label{eq:ZeroLagrangian}
        \partderiv{\boldsymbol{L}_t}{\bx}\biggl|_{(\bx_t^*, \bu_t^*)} = 0 \quad\text{and}\quad \partderiv{\boldsymbol{L}_t}{\bu}\biggl|_{(\bx_t^*, \bu_t^*)} = 0.
    \end{equation}
    The remaining three \ac{kkt} conditions are verified directly by Lemma~\ref{lem:StrongMaximumPrinciple} Condition~\ref{cond:WellPosedness} and \ref{cond:GreaterZero}. Hence, we have shown that the unique minimizer of $\overline{\boldsymbol{p}}_t(\bx_t,\bu_t)$ is given by the infinite-horizon optimal trajectory $(\bx^*, \bu^*)$ for all $t \in \N$.
    
    To derive a lower bound $\boldsymbol{\alpha}(\|(\bx_t\!-\!\bx_t^*,\! \bu_t\!-\!\bu_t^*)\|)$ for \eqref{eq:StrictDissipativity}, we introduce the shorthand $\bz_t := (\bx_t,\bu_t)$ and write
    \begin{equation*}
        \boldsymbol{L}_t(\bz_t,\mu_t) := \boldsymbol{L}_t(\bx_t,\bu_t,\mu_t).
    \end{equation*}
    Since $\boldsymbol{L}_t(\bz_t,\mu_t)$ is quadratic in $\bz_t$ by \eqref{eq:PotentialFunction}, its second-order Taylor expansion around $\bz_t^*$ is exact and reduces to
    \begin{alignat*}{2}
        \boldsymbol{L}_t(\bz_t,\mu_t) &= \overbrace{\boldsymbol{L}_t(\bz_t^*,\mu_t)}^{\stackrel{\ref{cond:GreaterZero}}{=}\,0} &&+\overbrace{\partderiv{\boldsymbol{L}_t}{\bz}\biggl|_{\rlap{$\scriptstyle(\bz_t^*,\mu_t)$}}}^{\smash{\stackrel{\raisebox{.8\height}{$\scriptscriptstyle\!\!\!\eqref{eq:ZeroLagrangian}\!\!\!$}}{=}\;0}} \quad (\bz_t - \bz_t^*) \\
        &&&+ \frac{1}{2}(\bz_t - \bz_t^*)^\top \partderivv{\boldsymbol{L}_t}{\bz} (\bz_t - \bz_t^*) \notag\\
        &\geq \rlap{$\displaystyle\frac{1}{2}\sigma_{\min}\biggl(\!\partderivv{\boldsymbol{L}_t}{\bz}\!\biggr) \|\bz_t - \bz_t^*\|^2$,}
    \end{alignat*}
    where the first term evaluates to zero due to Lemma~\ref{lem:StrongMaximumPrinciple} Condition~\ref{cond:GreaterZero}, the second one vanishes by \eqref{eq:ZeroLagrangian} and the last term is bounded from below by the smallest singular value of the Hessian. To make this lower bound uniform we define $\underline{\sigma}_{\min} := \inf_{t \in \N}\, \sigma_{\min}\!\bigl(\!\partderivv{\boldsymbol{L}_t}{\bz}\!\bigr)$ as the smallest singular value over all times $t \in \N$, which is positive by the uniform positive definiteness in Assumption~\ref{assum:UniquenessAndStrictConvexity}. We are left to show that $\overline{\boldsymbol{p}}_t(\bx_t,\bu_t)$ is lower bounded by the Lagrangian which follows from the observation that, even if we are not on the optimal trajectory $(\bx^*,\bu^*)$, $\mu_t^\top \boldsymbol{g}(\bx_t,\bu_t) \leq 0$ for all $(\bx_t,\bu_t) \in \bX \times \bU$ (see  Lemma~\ref{lem:StrongMaximumPrinciple}~Condition~\ref{cond:GreaterZero} and \eqref{eq:Constraints}), from which it follows that
    \begin{align*}
        \boldsymbol{L}_t(\bx_t,\bu_t,\mu_t) &= \overline{\boldsymbol{p}}_t(\bx_t,\bu_t) + \mu_t^\top \boldsymbol{g}(\bx_t,\bu_t)\\
        \frac{\underline{\sigma}_{\min}}{2} \|\bz_t - \bz_t^*\|^2&\leq \overline{\boldsymbol{p}}_t(\bx_t,\bu_t)\\
        \frac{\underline{\sigma}_{\min}}{2\eta_0} \|\bz_t - \bz_t^*\|^2 &\leq \lowtilde{\boldsymbol{p}}_t(\bx_t,\bu_t)
    \end{align*}
    where we used the fact that $\overline{\boldsymbol{p}}_t(\bx_t,\bu_t) := \eta_0\,\lowtilde{\boldsymbol{p}}_t(\bx_t,\bu_t)$.
    Hence,
    \begin{equation*}
        \boldsymbol{\alpha}(\|\bz_t - \bz_t^*\|) := \frac{\underline{\sigma}_{\min}}{2\eta_0}\|\bz_t - \bz_t^*\|^2
    \end{equation*}
    provides a uniform lower bound, which establishes the dissipation inequality \eqref{eq:StrictDissipativity} and completes the proof.
\end{pf}
With Theorem~\ref{thm:StrictDissipativity} we have shown that a time-varying \ac{pg} is strictly dissipative with respect to the infinite-horizon optimal trajectory $(\bx^*,\bu^*)$ obtained from solving \eqref{prob:InfCentralizedOCP} under strict convexity, \ac{ltv} cost and convex affine constraints (the input constraints may be non-affine). The following result is a direct conclusion from the fact that the rotated problem is dissipative with $\lowtilde{\boldsymbol{\Lambda}}_t(\bx_t) = 0$ (see \eqref{eq:RotatedStagePotentialFunction}).
\begin{cor}\label{cor:RotatedStrictDissipativity}
    Under Assumption~\ref{assum:UniquenessAndStrictConvexity}, \ref{assum:OptimalityCondition} and \ref{assum:SlaterlikeCondition}, Problem $(\lowtilde{\mathscr{P}}_N)$ is time-varying strictly dissipative with respect to the rotated stage potential function $\lowtilde{\boldsymbol{p}}_t(\bx_t,\bu_t)$ and the infinite-horizon optimal trajectory $(\bx^*, \bu^*)$ obtained from solving \eqref{prob:InfCentralizedOCP}.
\end{cor}

The next lemma is a consequence from the proof of Theorem~\ref{thm:StrictDissipativity} and is a stronger version of \cite[Assumption~8][]{grune_economic_2020}.
\begin{lem}\label{lem:ContinuousStorageFunction}
    Given that the assumptions of Theorem~\ref{thm:StrictDissipativity} hold, the respective storage function $\boldsymbol{\Lambda}_t(\bx_t)$ of \eqref{prob:CentralizedOCP} is continuous with $\boldsymbol{\gamma}_{\boldsymbol{\Lambda}}(\|\bx_1 - \bx_2\|) \in \mathcal{K}_\infty$ such that
    \vspace{-5pt}
    \begin{equation*}\vspace{-5pt}
        |\boldsymbol{\Lambda}_t(\bx_1) - \boldsymbol{\Lambda}_t(\bx_2)| \leq \boldsymbol{\gamma}_{\boldsymbol{\Lambda}}(\|\bx_1 - \bx_2\|)
    \end{equation*}
    holds for all $\bx_1, \bx_2 \in \bX$.
\end{lem}
\begin{pf}
    Starting from expression \eqref{eq:StorageFunction} we obtain
    \begin{align*}
        \!|\boldsymbol{\Lambda}_t(\bx_1) \!-\! \boldsymbol{\Lambda}_t(\bx_2)| &\leq\!\dfrac{\lowtilde{\vartheta}_{t\shortminus1\!}^\top}{\eta_0}\Bigl|\boldsymbol{g}_{\bx}\!\left(\boldsymbol{A}_{t\shortminus1}^{-1}(\bx_1 \!-\! \boldsymbol{B}_{t\shortminus1}\bu_{t\shortminus1}^*)\right)\notag\\
        &\qquad\qquad- \boldsymbol{g}_{\bx}\!\left(\boldsymbol{A}_{t\shortminus1}^{-1}(\bx_2 \!-\! \boldsymbol{B}_{t\shortminus1}\bu_{t\shortminus1}^*)\right)\Bigr|\!\\
        &\qquad\qquad\quad+\dfrac{\lambda_t^\top\!\!}{\eta_0\!\!}|\bx_1 - \bx_2|\notag\\
        &\leq\! \frac{\overline{\vartheta}\, l_{\boldsymbol{g}_{\bx}}}{\eta_0} \bigl\|\boldsymbol{A}_{t\shortminus1}^{-1}(\bx_1 \!-\!\bx_2)\bigr\| \!+\! \frac{\overline{\lambda}}{\eta_0} \|\bx_1 \!-\! \bx_2\|,\!\!
    \end{align*}
    where $l_{\boldsymbol{g}_{\bx}}$ is the (global) Lipschitz constant of $\boldsymbol{g}_{\bx}$, which exists since it is affine (Assumption~\ref{assum:OptimalityCondition} Condition~\ref{assum:Lipschitz}). Further, $\overline{\vartheta}$ and $\overline{\lambda}$ denote the uniform upper bounds on the multipliers from Assumption~\ref{assum:OptimalityCondition} Condition~\ref{assum:BoundedMultipliers}. Using $\|\boldsymbol{A}_{t\shortminus1}^{-1}\| \leq 1/\underline{\boldsymbol{\sigma}}_{\boldsymbol{A}}$, where $\underline{\boldsymbol{\sigma}}_{\boldsymbol{A}}$ denotes the smallest singular value of $\boldsymbol{A}_t$ uniformly over all $t \in \N$, we define
    \begin{equation}\label{eq:ContinuousStorageFunction}
        \boldsymbol{\gamma}_{\boldsymbol{\Lambda}}(\|\bx_1 - \bx_2\|) := \smash{\biggl(\frac{\overline{\vartheta}\, l_{\boldsymbol{g}_{\bx}}}{\eta_0\,\underline{\boldsymbol{\sigma}}_{\boldsymbol{A}}} + \frac{\overline{\lambda}}{\eta_0}\biggr)} \|\bx_1 - \bx_2\|.
    \end{equation}
\end{pf}\vspace{-20pt}
With this we have shown that the storage function associated to \eqref{prob:CentralizedOCP} is continuous using Lipschitz-type arguments. 

In order to show the turnpike property we additionally need a reachability condition of the infinite-horizon optimal trajectory $(\bx^*, \bu^*)$ \parencite[see][Assumption~1]{grune_strict_2018}.
\begin{assum}[Cheap reachability]\label{assum:CheapReachability}
    The infinite-horizon optimal trajectory $(\bx^*, \bu^*)$ is said to be \textit{cheaply reachable} if there exists a constant $E \in \domain{}$ such that the following inequality is satisfied
    \begin{equation}\label{eq:CheapReachability}
        \boldsymbol{V}_{\!\!N}(\bx_t,t) - \boldsymbol{p}_N(\bx^*,\bu^*, t) \leq E
    \end{equation}
    for all $t, N \in \N$ and $\bx_t \in \bX$.
\end{assum}
A stronger, more restrictive assumption that also guarantees such reachability is $n_{\bx}$-step reachability, where any feasible state can transition to the optimal trajectory within $n_{\bx}$ steps or fewer \parencite[cf.][]{weiss_controllability_1972}.

\begin{lem} \label{lem:Turnpike}
    Under Assumption~\ref{assum:UniquenessAndStrictConvexity}--\ref{assum:CheapReachability}, both Problem~\eqref{prob:CentralizedOCP} and its rotated counterpart $(\lowtilde{\mathscr{P}}_N)$ admit the turnpike property with respect to the infinite-horizon optimal trajectory $(\bx^*,\bu^*)$ as defined in Definition~\ref{def:Turnpike}.
\end{lem}

\begin{pf}
    Turnpike behaviour of Problem~\eqref{prob:CentralizedOCP} is a direct consequence of \cite[Theorem~1][]{grune_strict_2018}. To prove the same for the rotated problem $(\lowtilde{\mathscr{P}}_N)$ we first note that it is time-varying strictly dissipative with the storage function $\lowtilde{\boldsymbol{\Lambda}}_t(\bx_t) = 0$. We then further note that the rotated optimal value function can be upper bounded by\vspace{-5pt}
    \begin{align*}
        \lowtilde{\boldsymbol{V}}_{\!\!N}(\bx_t,t) &\leq \boldsymbol{V}_{\!\!N}(\bx_t,t) - \boldsymbol{p}_N(\bx^*,\bu^*, t)\\
        &\hspace{54pt} + \boldsymbol{\Lambda}_t(\bx_t) - \boldsymbol{\Lambda}_{t+N}(\bx_{t+N})\\
                                        &\stackrel{\!\eqref{eq:CheapReachability}\!}{\leq} E + \boldsymbol{\Lambda}_t(\bx_t) - \boldsymbol{\Lambda}_{t+N}(\bx_{t+N}).
    \end{align*}
    The rest of the proof follows analogously to the one of \cite[Theorem~1][]{grune_strict_2018}.
\end{pf}
With this we have shown the key properties needed for $\mathbb{P}$-practical asymptotic stability which are, time-varying strict dissipativity and the turnpike property. The remainder of this section is mostly bookkeeping that allows us to connect to the results of \cite{grune_economic_2020}.

The following result was given first in \cite{erdin_stability_2025}. We repeat it here for the sake of self-containedness.
\begin{lem}\label{lem:ContinuousStagePotential}
    If Assumption~\ref{assum:UniquenessAndStrictConvexity} is satisfied the stage potential function $\boldsymbol{p}_t(\bx_t,\!\bu_t)$ of Problem~\eqref{prob:CentralizedOCP} is continuous with $\boldsymbol{\gamma}_{\boldsymbol{p}} \in \mathcal{K}_{\infty}$ such that
    \begin{equation}\label{eq:ContinuousStagePotential}
        |\boldsymbol{p}_t(\bx_1,\!\bu_1) - \boldsymbol{p}_t(\bx_2,\!\bu_2)| \leq \boldsymbol{\gamma}_{\boldsymbol{p}}(\|(\bx_1 \!- \bx_2,\!\bu_1 \!- \bu_2)\|)
    \end{equation}
    holds for all $t \in \N$ and $(\bx_1,\!\bu_1), (\bx_2,\!\bu_2) \in \boldsymbol{\mathcal{Y}} \subseteq \bX \times \bU$ in a compact subset of the joint state and input space.
\end{lem}
\begin{pf}
    See Appendix \ref{app:ProofContinuousStagePotential}.
\end{pf}
The geometric interpretation of this result is that at any point $(\bx_2,\!\bu_2)$ we can fit a cone acting as an over-bound to $|\boldsymbol{p}_t(\bx_1,\!\bu_1) - \boldsymbol{p}_t(\bx_2,\!\bu_2)|$ given that $(\bx_1,\!\bu_1)$ stays in a compact set around $(\bx_2,\!\bu_2)$.

The next lemma is a consequence from the continuity of the stage potential function and the storage function.
\begin{lem}\label{lem:ContinuousRotatedStagePotential}
    Given that Assumption~\ref{assum:UniquenessAndStrictConvexity}, \ref{assum:OptimalityCondition} and \ref{assum:SlaterlikeCondition} hold, the rotated stage potential function $\lowtilde{\boldsymbol{p}}_t(\bx_t,\!\bu_t)$ is continuous with $\boldsymbol{\gamma}_{\lowtilde{\boldsymbol{p}}} \in \mathcal{K}_{\infty}$ such that
    \begin{equation} \label{eq:ContinuousRotatedStagePotential}
        |\lowtilde{\boldsymbol{p}}_t(\bx_1,\!\bu_1) - \lowtilde{\boldsymbol{p}}_t(\bx_2,\!\bu_2)| \leq \boldsymbol{\gamma}_{\lowtilde{\boldsymbol{p}}}(\|(\bx_1 \!- \bx_2,\bu_1 \!- \bu_2)\|)
    \end{equation}
    holds for all $t \in \N$ and $(\bx_1,\!\bu_1), (\bx_2,\!\bu_2) \in \boldsymbol{\mathcal{Y}} \subseteq \bX \times \bU$ in a compact subset of the joint state and input space.
\end{lem}
\begin{pf}
    Expanding the left-hand side of \eqref{eq:ContinuousRotatedStagePotential} and applying Lemma~\ref{lem:ContinuousStorageFunction} and \ref{lem:ContinuousStagePotential}, yields
    \begin{align*}
        \rlap{$\hspace{-5pt}|\lowtilde{\boldsymbol{p}}_t(\bx_1,\!\bu_2) - \lowtilde{\boldsymbol{p}}_t(\bx_2,\!\bu_2)|$}\qquad& \notag\\
        &\leq |\boldsymbol{p}_t(\bx_1,\!\bu_1) \!-\! \boldsymbol{p}_t(\bx_2,\!\bu_2)| + |\boldsymbol{\Lambda}_t(\bx_1) \!-\! \boldsymbol{\Lambda}_t(\bx_2)| \\
        &\quad+ |\boldsymbol{\Lambda}_{t+1}(\boldsymbol{A}_t \bx_1 \!+\! \boldsymbol{B}_t \bu_1) \!-\! \boldsymbol{\Lambda}_{t+1}(\boldsymbol{A}_t \bx_2 \!+\! \boldsymbol{B}_t \bu_2)| \notag\\
        &\stackrel{\!\!\!\!\!\!\!\!\eqref{eq:ContinuousStagePotential}, \eqref{eq:ContinuousStorageFunction}\!\!\!\!\!\!\!\!}{\leq}\! \boldsymbol{\gamma}_{\boldsymbol{p}}(\|(\bx_1 \!- \bx_2,\bu_1 \!- \bu_2)\|) + \boldsymbol{\gamma}_{\boldsymbol{\Lambda}}(\|\bx_1 - \bx_2\|) \notag\\
        &\quad + (\overline{\boldsymbol{\sigma}}_{\!\boldsymbol{A}} + \overline{\boldsymbol{\sigma}}_{\!\boldsymbol{B}})\,\boldsymbol{\gamma}_{\boldsymbol{\Lambda}}(\|(\bx_1 - \bx_2, \bu_1 - \bu_2)\|),
    \end{align*}
    where $\overline{\boldsymbol{\sigma}}_{\!\boldsymbol{A}}$ and $\overline{\boldsymbol{\sigma}}_{\!\boldsymbol{B}}$ denote the largest singular value of the state transition matrix $\boldsymbol{A}_t$ and the input matrix $\boldsymbol{B}_t$, respectively, uniformly over all $t \in \N$.
\end{pf}
The following corollary immediately follows from the circumstance that $\lowtilde{\boldsymbol{p}}_t(\bx_t^*, \bu_t^*) = 0$.
\begin{cor}\label{cor:UpperBoundedRotatedStagePotential}
    Suppose Assumption~\ref{assum:UniquenessAndStrictConvexity}, \ref{assum:OptimalityCondition} and \ref{assum:SlaterlikeCondition} hold. Then the rotated stage potential function $\lowtilde{\boldsymbol{p}}_t(\bx_t, \bu_t)$ is upper bounded by $\boldsymbol{\gamma}_{\lowtilde{\boldsymbol{p}}} \in \mathcal{K}_{\infty}$ such that
    \begin{equation*}\label{eq:UpperBoundedRotatedStagePotential}
        \lowtilde{\boldsymbol{p}}_t(\bx_t, \bu_t) \leq \boldsymbol{\gamma}_{\lowtilde{\boldsymbol{p}}} (\|(\bx_t - \bx_t^*, \bu_t - \bu_t^*)\|),
    \end{equation*}
    holds for all $t \in \N$ and $(\bx_t, \bu_t) \in \boldsymbol{\mathcal{B}}_{\delta_{\tilde{p}}}(\bx_t^*, \bu_t^*)$, with $\boldsymbol{\mathcal{B}}_{\delta_{\tilde{p}}}(\bx_t^*, \bu_t^*)$ denoting a neighbourhood around the infinite-horizon optimal trajectory.
\end{cor}

\begin{figure}[t]
    \centering
    \includegraphics[width=230pt]{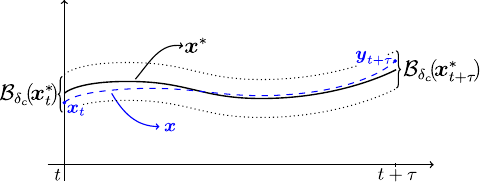}
    \caption{Local controllability of trajectory $\bx$ around the optimal trajectory $\bx^*$ (Figure adapted from \cite[Figure~5.1][]{pirkelmann_economic_2020}).}
    \label{fig:LocalControllability}
\end{figure}

The last result needed to connect to the work of \cite{grune_economic_2020} is continuity of the optimal value functions $\boldsymbol{V}_{\!\!N}(\bx_t,t)$ and $\lowtilde{\boldsymbol{V}}_{\!\!N}(\bx_t,t)$, for \eqref{prob:CentralizedOCP} and $(\lowtilde{\mathscr{P}}_N)$ respectively. To establish this, we introduce one final assumption, which is local controllability.
\begin{assum}[Local controllability] \label{assum:LocalControllability}
    Suppose system \eqref{eq:Dynamics} is \textit{locally controllable} along the trajectory pair $(\bx^*, \bu^*)$. That is, there exist a time horizon $\tau \in \N$, a constant $\delta_c > 0$, and class-$\mathcal{K}_\infty$ functions $\boldsymbol{\gamma}_x(\delta), \boldsymbol{\gamma}_u(\delta), \boldsymbol{\gamma}_c(\delta)$ such that the following holds for all $t \in \N$.
    For any $\bx_t \!\in \mathcal{B}_{\delta_c}(\bx_t^*)$ and $\boldsymbol{y}_{t+\tau} \!\in \mathcal{B}_{\delta_c}(\bx_{t+\tau}^*)$, there exists an input sequence $\{\bu_k\}_{k=t}^{t+\tau} \in \bU$ steering the system from $\bx_t$ to $\bx_{t+\tau} = \boldsymbol{y}_{t+\tau}$, such that
    \begin{align*}
        \|\bx_{t+k} - \bx_{t+k}^*\| &\leq \boldsymbol{\gamma}_x(\boldsymbol{\delta}), \\
        \|\bu_{t+k} - \bu_{t+k}^*\| &\leq \boldsymbol{\gamma}_u(\boldsymbol{\delta}), \\
        |\hat{\boldsymbol{p}}_t(\bx_{t+k}, \bu_{t+k})| &\leq \boldsymbol{\gamma}_c(\boldsymbol{\delta})
    \end{align*}
    holds for all $k \in \{t,\ldots,t+\tau\}$, with $\boldsymbol{\delta} := \max\{\|\bx_t-\bx_t^*\|,\|\boldsymbol{y}_{t+\tau} -\bx_{t+\tau}^*\|\}$.
\end{assum}
This simply means that any two points $(\bx_t, \boldsymbol{y}_{t+\tau})$ within a $\delta_c$-vicinity of the optimal trajectory can be connected by a sub-optimal trajectory in forward time. For a graphical interpretation of this assumption, see Figure~\ref{fig:LocalControllability}.
\FloatBarrier

Given local controllability and the aforementioned assumptions, the following lemma is an immediate consequence of \cite[Theorem~2][]{grune_strict_2018}.
\begin{lem}\label{lem:ContinuousValueFunction}
    Let Assumptions~\ref{assum:UniquenessAndStrictConvexity}--\ref{assum:LocalControllability} hold. Then, the optimal value function of Problem~\eqref{prob:CentralizedOCP} and $(\lowtilde{\mathscr{P}}_N)$ is (approximately) continuous around the infinite-horizon optimal trajectory $\bx^*$, in the sense that, for all $t \in \N$, $N \in \N \cup \{\infty\}$ and all $\bx_t \in \boldsymbol{\mathcal{B}}_{\delta_{\tilde{P}}}(\bx_t^*)$, the following holds
    \begin{subequations}
    \begin{align}
        |\boldsymbol{V}_N(\bx_t,t) - \boldsymbol{V}_N(\bx_t^*,t)| &\leq \boldsymbol{\gamma}_{\boldsymbol{V}}(N,\|\bx_t - \bx_t^*\|),\\
        |\lowtilde{\boldsymbol{V}}_N(\bx_t,t) - \lowtilde{\boldsymbol{V}}_N(\bx_t^*,t)| &\leq \boldsymbol{\gamma}_{\lowtilde{\boldsymbol{V}}}(\|\bx_t - \bx_t^*\|).\label{eq:ContinuousRotatedOptimalValueFunction} 
    \end{align}
    \end{subequations}
    Here $\boldsymbol{\gamma}_{\boldsymbol{V}}(N,\Delta\bx)$ ($\Delta\bx:= \|\bx_t - \bx_t^*\|$) is a function that is monotonically decreasing in $N$ for fixed $\Delta\bx$ and monotonically increasing in $\Delta\bx$ for fixed $N$. Additionally, in the limit $\boldsymbol{\gamma}_{\boldsymbol{V}}(\infty,0) = 0$, $\boldsymbol{\gamma}_{\boldsymbol{V}}(\infty,\Delta\bx) \in \mathcal{K}_{\infty}$ and $\boldsymbol{\gamma}_{\lowtilde{\boldsymbol{V}}}$ is defined analogously to $\boldsymbol{\gamma}_{\boldsymbol{V}}(N,\bullet)$, i.e. it is monotonically increasing in $\Delta\bx$.
\end{lem}

\begin{pf}
     Under Assumption~\ref{assum:UniquenessAndStrictConvexity}--\ref{assum:LocalControllability}, together with Lemma~\ref{lem:ContinuousStagePotential}, \ref{lem:ContinuousRotatedStagePotential} and Corollary~\ref{cor:UpperBoundedRotatedStagePotential}, Problem~\eqref{prob:CentralizedOCP} and $(\lowtilde{\mathscr{P}}_N)$ satisfy all assumptions of \cite[Theorem~2][]{grune_strict_2018} from which the following upper bounds follow
     \begin{equation*}
         \boldsymbol{\gamma}_{\boldsymbol{V}}(N,\|\bx_t - \bx_t^*\|), \qquad \boldsymbol{\gamma}_{\lowtilde{\boldsymbol{V}}}(N,\|\bx_t - \bx_t^*\|).
     \end{equation*}
     The upper bound for \eqref{eq:ContinuousRotatedOptimalValueFunction} is then simply obtained by setting $N=1$
     \begin{equation*}
         \boldsymbol{\gamma}_{\lowtilde{\boldsymbol{V}}}(\|\bx_t - \bx_t^*\|) := \boldsymbol{\gamma}_{\lowtilde{\boldsymbol{V}}}(1,\|\bx_t - \bx_t^*\|).
     \end{equation*}
\end{pf}\vspace{-20pt}
Note that $\boldsymbol{\gamma}_{\boldsymbol{V}}(N,0)$, $\boldsymbol{\gamma}_{\lowtilde{\boldsymbol{V}}}(0)$ is not necessarily zero which is why this condition is referred to as approximate continuity. Only in the limit it holds that $\lim_{N \rightarrow \infty} \boldsymbol{\gamma}_{\boldsymbol{V}}(N,0) = 0$.

With Lemma~\ref{lem:ContinuousValueFunction}, we complete the connection between our assumptions and those of \cite{grune_economic_2020}. Assumptions~\ref{assum:UniquenessAndStrictConvexity}--\ref{assum:LocalControllability} recover their assumptions exactly, with the exception of Lemma~\ref{lem:ContinuousValueFunction} and Corollary~\ref{cor:UpperBoundedRotatedStagePotential}, which are relaxed versions of \cite[Assumption~7~and~6][]{grune_economic_2020}, respectively. Under compact constraints these relaxed versions are in fact equivalent to the originals, which lets us invoke the stability results of \cite{grune_economic_2020} and thereby obtain $\mathbb{P}$-practical asymptotic convergence to the infinite-horizon optimal trajectory $(\bx^*,\bu^*)$, with the $\mathbb{P}$-set shrinking to zero as the horizon $N \to \infty$.

In short, under Assumptions~\ref{assum:OptimalityCondition}--\ref{assum:LocalControllability}, Problem~\eqref{prob:CentralizedOCP} satisfies all the assumptions of \cite{grune_economic_2020}.

\begin{rem} \label{rem:RelaxedAssumption}
    Since the two assumptions we relax \parencite[Assumptions~6~and~7][]{grune_economic_2020} are used only locally, around the infinite-horizon optimal trajectory $(\bx^*,\bu^*)$, we conjecture that our results extend to also hold without relying on compact constraints.
\end{rem}

To summarize, we analysed the stability of Problem~\eqref{prob:GNEP} via its centralized reformulation \eqref{prob:CentralizedOCP}. We provided a new set of sufficient conditions under which the constrained Problem~\eqref{prob:CentralizedOCP} is time-varying strictly dissipative with respect to the infinite-horizon optimal trajectory $(\bx^*,\bu^*)$. This result is relevant not only to the game-theoretic community but also to the \ac{mpc} community, since our \ac{ltv} setting with quadratic objective and convex constraints extends existing results for time-invariant \ac{mpc}. We related our assumptions to those of \cite{grune_economic_2020}, drawing primarily on \cite{grune_strict_2018}, to establish $\mathbb{P}$-practical asymptotic stability. Notably, we required no additional constraints or terminal penalties on the multi-agent game \eqref{prob:GNEP}, relying solely on the structure of Problem~\eqref{prob:GNEP} itself.

%% file: figures/problems.tex
\newdimen\boxwidth\boxwidth=.42\linewidth
\noindent
\begin{tikzpicture}[%
    dashedBox/.style={box, minimum height=2.8cm, dashed, draw=gray!90},
    contentBox/.style={box, minimum width=\boxwidth-2\mgn, minimum height=1cm, outer sep=\mgn}
    ]
    \node[dashedBox, anchor=west, minimum width=\boxwidth, label={[labelStyle]north west:GNEPs}] (gneps) at (1\mgn, 0) {};
    \node[contentBox, anchor=north] (gnep) at (gneps.north) {\eqref{prob:InfGNEP} \rlap{$\forall v \in \V$}};
    \node[contentBox, anchor=south] (rhg) at (gneps.south) {\eqref{prob:GNEP} \rlap{$\forall v \in \V$}};
    
    \node[dashedBox, anchor=east, minimum width=\boxwidth, label={[labelStyle]north west:PGs}] (pgs) at (\linewidth-1\mgn, 0) {};
    \node[contentBox, anchor=north] (pg) at (pgs.north) {\eqref{prob:InfCentralizedOCP}};
    \node[contentBox, anchor=south] (rhpg) at (pgs.south) {\eqref{prob:CentralizedOCP}};

    \draw[arrow]
        (gnep.south)edge (rhg.north)
        (pg.south)  to (rhpg.north);
    \draw[arrow, latex-latex]
        (gnep.east) edge (pg.west)
        (rhg.east)  to (rhpg.west);

    \path (gneps.west) -- node[fill=white, inner sep=0pt, font=\bf\scriptsize, rotate=90, align=center] {Variational\\ Equilibrium} (pgs.east);
\end{tikzpicture}

%% file: figures/kkt.tex
\noindent
\begin{tikzpicture}[dashedBox/.style={box, minimum height=1.8cm, dashed, draw=gray!90}]
    \node[dashedBox, anchor=west, minimum width=.39\linewidth, label={[labelStyle]north west:Primal Problem}] (primal) at (0, 0) {$
    \begin{alignedat}{2}
        \min_{\bx_t,\bu_t}\; & \overline{\boldsymbol{p}}_t(\bx_t,\bu_t) \\
       \text{s.t.} \;& (\bx_t, \bu_t) \in \bX \times \bU
    \end{alignedat}$};
    \node[dashedBox, anchor=east, minimum width=.37\linewidth, label={[labelStyle]north west:Dual Problem}] (dual) at (\linewidth-6\mgn, 0) {$
    \begin{alignedat}{2}
        \min_{\bx_t,\bu_t,\mu_t}  & \boldsymbol{L}_t(\bx_t,\bu_t,\mu_t) \\
            \text{s.t.}\;\,{} &\mu_t \geq 0
    \end{alignedat}
    $};

    \draw[latex-latex] (primal.east) to node[align=center, font=\bf\scriptsize, fill=white, inner sep=0pt, text width=2cm, rotate=90] {Strict\\ Convexity}(dual.west);
\end{tikzpicture}

%% file: sections/03_examples.tex
\section{Simulation Study}\label{sec:SimulationStudy}

The example studied here was originally proposed by \cite{hall_receding_2022} and therein analysed for the time-invariant case. We consider an electric grid with 10 prosumers\footnote{A prosumer is a consumer that does not just consume energy but can also produce and store it.} that each have the dynamics
\begin{equation*}
x_{t+1}^v =
        \begin{bmatrix}
            1 & 0 \\
            0 & \alpha_t^v
        \end{bmatrix}
        x_t^v +
        \begin{bmatrix}
            1 & 0 \\
            0 & \beta_t^v
        \end{bmatrix}
        u_t^v +
        \begin{bmatrix}
            \shortminus e_t{}^{v,\text{ref }} \\
            0
        \end{bmatrix},
\end{equation*}
where the state vector $x_t^v := (\zeta_t^v, q_t^v)$ collects energy debt $\zeta_t^v$ and the battery state of charge $q_t^v$, and the input vector $u_t^v := (e_t^v, s_t^v)$ collects energy consumption $e_t^v$ and battery charging/discharging inputs $s_t^v$. Further $e_t{}^{v,\text{ref }}$ denotes the desired reference energy consumption of each household. Each prosumer $v$ is minimizing the cost
\vspace{-10pt}
\begin{multline*}
    \ell_t^v(x_t^v,\!\bu_t) := \bigl(x_t^v\bigr){}^{\!\top}\!
        \begin{bmatrix}
            \gamma_{1,t}^v & 0 \\
            0              & \gamma_{2,t}^v
         \end{bmatrix}
         \!x_t^v \\
         + \sum_{i \in \V}\Biggl(\bigl(u_t^v\bigr){}^{\!\top}\!
         \begin{bmatrix}
             \rho_{1,t}^v & \rho_{1,t}^v \\
             \rho_{1,t}^v & \rho_{1,t}^v
         \end{bmatrix}
         \!u_t^i \Biggr) + c_t^v u_t^v,
\end{multline*}
which introduces a coupling through the input cost terms. An additional coupling is introduced through the box constraints, defined for all $\Theta \in \{\zeta_t^v, q_t^v, e_t^v, s_t^v, e_t^v+s_t^v, \boldsymbol{L}_t\}$, such that
\begin{equation*}
    \Theta \in \bigl[\,\underline{\Theta},\; \overline{\Theta}\,\bigr]
\end{equation*}
holds for all $t \in \N$. Specifically, it is introduced through the aggregative load $\boldsymbol{L}_t$, which is defined as
\begin{equation*}
    \boldsymbol{L}_t := \smash{\sum_{v \in \V}} (e_t^v + s_t^v - g_t^v) + \smash{\sum_{v \in \mathcal{P}}} e_t^v,
\end{equation*}
where $\mathcal{P}$ denotes the set of passive consumers that do not participate in the demand-side management scheme and $g_t^v$ denotes energy produced by renewable energy sources.

We implemented the centralized algorithm \eqref{prob:CentralizedOCP} with the above dynamics and cost as well as the parameters from \cite{hall_receding_2022} using \texttt{CVXOPT} \parencite{diamond_cvxpy_2016,agrawal_rewriting_2018} in \texttt{Python}\footnote{The code and all the parameters used can be found here: \url{https://github.com/Gedlex/time-varying-RHGs.git}}. The system is simulated for a duration of 2 days and the infinite-horizon optimal trajectory is approximated by solving the open-loop problem over a horizon of 24 days.

\begin{figure}[ht]
    \centering
    \includegraphics[width=3.32in]{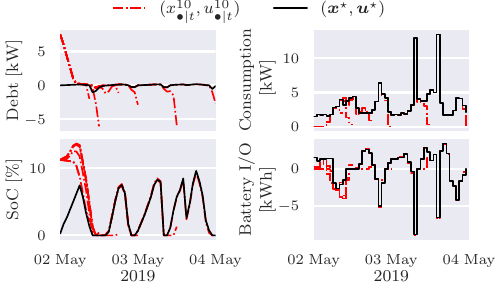}\vspace{-5pt}
    \caption{Open-loop trajectories of agent 10 for increasing horizon lengths $N \in \{6,\, 9,\, 12,\, 18,\, 24,\, 36,\, 48\}$.}
    \label{fig:OpenLoopTrajectories}
\end{figure}

The open-loop behaviour is exemplarily depicted for household 10 (Figure \ref{fig:OpenLoopTrajectories}) for prediction horizons of increasing length. In the obtained predictions we can very nicely observe the turnpike property predicted by Lemma \ref{lem:Turnpike}.

\begin{figure}[ht]
    \centering
    \includegraphics[width=3.32in]{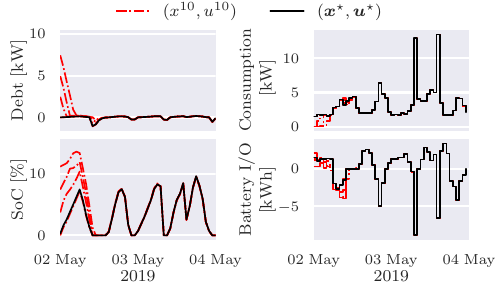}\vspace{-5pt}
    \caption{Closed-loop trajectories of agent 10 for varying initial conditions with horizon length $N = 24$.}
    \label{fig:ClosedLoopTrajectories}
\end{figure}

Figure \ref{fig:ClosedLoopTrajectories} illustrates how the closed-loop trajectories of household 10 converge to the infinite-horizon optimal trajectory given different initial conditions. The simulation hereby showed that all the agents converge to the optimal solution within roughly 36 hours.
\nopagebreak

%% file: sections/98_conclusion.tex
\FloatBarrier
\section{Conclusion}\label{sec:Conclusion}
In this paper we have presented stability guarantees for \aclp{rhg} with a \acl{pg} structure, a time-varying quadratic objective and \ac{ltv} dynamics. We have shown under which conditions a time-varying quadratic multi-agent objective admits a potential function and have used the centralized reformulation provided by the \ac{pg} structure to analyse the stability of the game with respect to the \ac{vgne} of \eqref{prob:InfGNEP}. Thereby, we have derived a new optimality condition, which together with a Slater-like condition yields time-varying strict dissipativity with respect to the infinite-horizon optimal trajectory $(\bx^*,\bu^*)$ (or \ac{vgne}). Lastly we have established a rigorous connection to existing \ac{empc} literature verifying $\mathbb{P}$-practical asymptotic stability.

Future research might be concerned with showing dissipativity without relying on strong duality, i.e. for problems that don't satisfy the optimality condition (Assumption \ref{assum:OptimalityCondition}) and to extend these results beyond \ac{ltv} dynamics to more general nonlinear time-varying systems. A generalisation of these results to the non-potential setting seems particularly promising since it covers a much richer class of problem formulations that is not restricted to the analysis of \acp{vgne}. Recent results \parencite{hall_stability_2024,hall_system-theoretic_2026} achieve this for the time-invariant case using arguments from monotone-operator theory while the latter extends tools such as dissipativity and turnpike to the game-theoretic setting.

%% file: sections/99_appendix.tex
\appendix
\section{Proof of Lemma \ref{lem:ContinuousStagePotential}} \label{app:ProofContinuousStagePotential}

Let us first introduce the shorthand notation
\begin{equation*}
    \boldsymbol{p}_t(\bz_t) := \bz_t^\top \boldsymbol{M}_t \bz_t + \boldsymbol{m}_t \bz_t,
\end{equation*}
with $\bz_t := (\bx_t,\bu_t)$ being the combined state and input and the stacked quantities $\boldsymbol{M}_t$ and $\boldsymbol{m}_t$ being defined as
\begin{equation*}
    \boldsymbol{M}_t := \blkdiag( \boldsymbol{Q}_t, \boldsymbol{R}_t),\quad \boldsymbol{m}_t := [\boldsymbol{q}_t\;\,  \boldsymbol{r}_t].
\end{equation*}
To derive the over-bound $\boldsymbol{\gamma}_{\boldsymbol{p}}(\|(\bx_1 \!- \bx_2,\!\bu_1 \!- \bu_2)\|)$ we expand the expression
\begin{align*}
    \!\!\!\boldsymbol{p}_t(\bz_1) \!-\! \boldsymbol{p}_t(\bz_2) \!=\,& \bz_1^{\top\!} \boldsymbol{M}_t \bz_1 \!-\! \bz_2^{\top\!} \boldsymbol{M}_t \bz_2 \!+\! \boldsymbol{m}_t (\bz_1 \!-\! \bz_2)\!\!\!\\
    \!=\,& \bz_1^{\top\!} \boldsymbol{M}_t \bz_1 \!-\! \bz_1^{\top\!} \boldsymbol{M}_t \bz_2 \!+\! \bz_1^{\top\!} \boldsymbol{M}_t \bz_2 \notag\\
    & \hspace{36pt}-\! \bz_2^{\top\!} \boldsymbol{M}_t \bz_2 \!+\! \boldsymbol{m}_t (\bz_1 \!-\! \bz_2)\!\!\! \\
    \!=\,& \bz_1^{\top\!} \boldsymbol{M}_t (\bz_1 \!-\! \bz_2) \!+\! (\bz_1 \!-\! \bz_2)^{\top\!} \boldsymbol{M}_t \bz_2 \notag\\
    & \;+\! \boldsymbol{m}_t (\bz_1 \!-\! \bz_2)
\end{align*}
from which we then take the absolute value to derive the desired over-bound
\begin{align*}
    |\boldsymbol{p}_t(\bz_1) \!-\! \boldsymbol{p}_t(\bz_2)| \stackrel{\!\smash{\text{(I)}}\!}{\leq}\;& |\bz_1^{\top\!} \boldsymbol{M}_t (\bz_1 \!-\! \bz_2)| \!+\! |(\bz_1 \!-\! \bz_2)^{\top\!} \boldsymbol{M}_t \bz_2| \notag\\
    & \;+\! |\boldsymbol{m}_t (\bz_1 \!-\! \bz_2)| \\
    \stackrel{\!\!\smash{\text{(II)}}\!\!}{\leq}\;& \|\boldsymbol{M}_t\| \bigl(\|\bz_1\| \!+\! \|\bz_2\|\bigr) \cdot \|\bz_1 \!-\! \bz_2\| \notag\\
    & \hspace{49pt}+ \|\boldsymbol{m}_t\| \cdot \|\bz_1 \!-\! \bz_2\|, \!\!\!
\end{align*}
where we first apply the triangle inequality in (I), followed by the induced matrix norm combined with the Cauchy-Schwarz inequality in (II). An upper bound for this expression is given by
\begin{equation*}
    \boldsymbol{\gamma}_{\boldsymbol{p}}(\|\bz_1 - \bz_2\|) := (2\,\overline{\boldsymbol{\sigma}}_{\bz} \overline{\bz} + \overline{\boldsymbol{m}}) \|\bz_1 - \bz_2\|,
\end{equation*}
with $\overline{\boldsymbol{\sigma}}_{\bz} := \sup_{t \in \N} \max(\overline{\boldsymbol{\sigma}}_{\boldsymbol{Q}_t}, \overline{\boldsymbol{\sigma}}_{\boldsymbol{R}_t})$ denoting the maximum singular value of $\boldsymbol{Q}_t$, $\boldsymbol{R}_t$ uniformly over all $t \in \N$ and $\overline{\boldsymbol{m}} := \sup_{t \in \N} \max(\overline{\boldsymbol{q}}_t, \overline{\boldsymbol{r}}_t)$ defined analogously. $\overline{\bz} := \max_{\bz \;\in\; \boldsymbol{\mathcal{Y}}} \|\bz\|$ denotes the maximum value of $\bz \in \boldsymbol{\mathcal{Y}}$, which is bounded since $\boldsymbol{\mathcal{Y}}$ is compact.\hfill$\blacksquare$\par\null